%% file: conference_101719.tex
\documentclass[conference]{IEEEtran}
\IEEEoverridecommandlockouts
\usepackage{cite}
\usepackage{amsmath,amssymb,amsfonts}
\usepackage{graphicx}
\usepackage{textcomp}
\usepackage{xcolor}
\usepackage{adjustbox}
\usepackage{algorithm}
\usepackage[]{algpseudocode}
\usepackage{color,colortbl}
\usepackage{enumitem}
\usepackage{multirow}
\usepackage{pgfplots}	
\usepgfplotslibrary{statistics}
\pgfplotsset{compat = 1.16}
\usepackage{tikz}
\usetikzlibrary{matrix,shapes,arrows,positioning,chains, calc, automata}
\usepackage{todonotes} 
\usepackage{url}

\newcommand\copyrightnotice[1]{
	\begin{tikzpicture}[remember picture,overlay]
		\node[anchor=south,yshift=10pt] at (current page.south) {\fbox{\parbox{\dimexpr\textwidth-\fboxsep-\fboxrule\relax}{#1}}};
	\end{tikzpicture}
}

\include{symbols}

\include{settings}

\def\BibTeX{{\rm B\kern-.05em{\sc i\kern-.025em b}\kern-.08em
    T\kern-.1667em\lower.7ex\hbox{E}\kern-.125emX}}
\begin{document}

\title{Solving the Kidney Exchange Problem Using Privacy-Preserving Integer Programming\\\LARGE (Updated and Extended Version)
\thanks{This work was funded by the Deutsche Forschungsgemeinschaft (DFG, German Resarch Foundation) - project number (419340256) and NSF grant CCF-1646999. Any opinion, findings, and conclusions or recommendations expressed in this material are those of the author(s) and do not necessarily reflect the views of the National Science Foundation.}}

\author{\IEEEauthorblockN{Malte Breuer$ ^\ast $, Pascal Hein$ ^\dagger $, Leonardo Pompe$ ^\dagger $, Ben Temme$ ^\dagger $, Ulrike Meyer$ ^\ast $, Susanne Wetzel$ ^\ddagger $}
	\IEEEauthorblockA{$ ^\ast $ RWTH Aachen University, Aachen, Germany: \{breuer, meyer\}@itsec.rwth-aachen.de} 
	\IEEEauthorblockA{$ ^\dagger $ RWTH Aachen University, Aachen, Germany: \{pascal.hein, leonardo.pompe, ben.temme\}@rwth-aachen.de} 
	\IEEEauthorblockA{$ ^\ddagger $ Stevens Institute of Technology, Hoboken, NJ, USA: swetzel@stevens.edu} 
}

\maketitle

\copyrightnotice{\copyright\space 2022 IEEE. This is the updated and extended version of the work published in \emph{19th Annual International Conference on Privacy, Security and Trust (PST2022), August 22-24, 2022, Fredericton, Canada / Virtual Conference}, https://doi.org/10.1109/PST55820.2022.9851968. It is posted here for your personal use. Not for redistribution.}

\begin{abstract}
\input{chapters/abstract}
\end{abstract}

\input{chapters/introduction}
\input{chapters/preliminaries}
\input{chapters/branch_and_bound}
\input{chapters/protocol}
\input{chapters/evaluation}

\input{chapters/relatedWork}
\input{chapters/conclusion}

\section*{Source Code}
The MP-SPDZ source code of our protocol~$ \kepIpProtocol $ (cf.~Protocol~\ref{prot:kep_ip}) and a script for executing the protocol on a sample set of randomly generated patient-donor pairs are available under the following link:
\url{https://gitlab.com/rwth-itsec/kep-ip}

While we cannot publish the inputs from our real-world data set, this data can be requested from UNOS by anyone who is interested. 

\bibliographystyle{IEEEtranS}
\bibliography{references}

\appendix
\input{chapters/appendix}

\end{document}

%% file: symbols.tex
\DeclareMathAlphabet{\pazocal}{OMS}{zplm}{m}{n}

\newcommand*{\customParagraph}[1]{
	\medskip 
	
	\noindent \textbf{#1}}

\newcommand*{\numCompPeers}{\ensuremath{\kappa}}

\newcommand*{\breuerSecSharingProt}{\ensuremath{\small{\textsc{Kep-Rnd-SS}}}}

\newcommand*{\compGraphMatrix}{\ensuremath{M}}

\newcommand*{\simplexFrequency}{\ensuremath{\lambda}}
\newcommand*{\babFrequency}{\ensuremath{\kappa}}

\newcommand*{\nodes}{\ensuremath{V}}
\newcommand*{\numColumns}{\ensuremath{m}}
\newcommand*{\numConstraints}{\ensuremath{m}}
\newcommand*{\numVariables}{\ensuremath{n}}
\newcommand*{\objectiveVec}{\ensuremath{c}}
\newcommand*{\boundVec}{\ensuremath{b}}
\newcommand*{\constraintMatrix}{\ensuremath{A}}
\newcommand*{\constraintEntry}{\ensuremath{a}}
\newcommand*{\objectiveFunc}{\ensuremath{x \mapsto \objectiveVec^\top x}}
\newcommand*{\solutionVec}{\ensuremath{x}}
\newcommand*{\objectiveValue}{\ensuremath{z}}
\newcommand*{\slackVector}{\ensuremath{s}}
\newcommand*{\cycleBound}{\ensuremath{l}}
\newcommand*{\incumbent}{\ensuremath{X^*}}
\newcommand*{\incumbentValue}{\ensuremath{z^*}}
\newcommand*{\simplexSolutionIncomplete}{\ensuremath{X'}}
\newcommand*{\simplexSolutionComplete}{\ensuremath{X}}
\newcommand*{\simplexSolutionValue}{\ensuremath{z}}

\newcommand*{\denominator}{\ensuremath{d}}
\newcommand*{\upperbound}{\ensuremath{\overline{b}}}
\newcommand*{\explicitSubsetVar}[1]{\ensuremath{x_{#1}}}
\newcommand*{\tableau}{\ensuremath{T}}
\newcommand*{\initialTableau}{\ensuremath{\tableau}}
\newcommand*{\partialSolutionVector}{\ensuremath{Q}}
\newcommand*{\newConstraintVector}{\ensuremath{F}}

\newcommand*{\solutionMatrix}{\ensuremath{A}}
\newcommand*{\outputDonors}{\ensuremath{D}}
\newcommand*{\outputRecipients}{\ensuremath{R}}
\newcommand*{\singleCycle}{\ensuremath{C}}
\newcommand*{\cycles}{\ensuremath{\mathcal{\singleCycle}}}
\newcommand*{\cycleVariable}{\ensuremath{x_\singleCycle}}
\newcommand*{\singleSubset}{\ensuremath{S}}
\newcommand*{\subsets}{\ensuremath{\mathcal{\singleSubset}}}
\newcommand*{\subsetVariable}{\ensuremath{x_\singleSubset}}
\newcommand*{\subsetVariableIndexed}[1]{\ensuremath{x_{\singleSubset_{#1}}}}
\newcommand*{\subsetMap}{\ensuremath{\mu}}
\newcommand*{\secretSubsetMap}{\ensuremath{\mathcal{M}}}
\newcommand*{\completeGraph}{\ensuremath{K}}
\newcommand*{\completeGraphSize}{\ensuremath{k_i}}
\newcommand*{\maxCompleteGraphSize}{\ensuremath{k_{\text{max}}}}

\newcommand*{\largestValuePossible}{\ensuremath{\numparties \sharedMult \enc{\denominator}}}

\newcommand*{\branchAndBoundStructure}{\ensuremath{\mathcal{B}}}
\newcommand*{\numSimplexIterations}{\ensuremath{\mathcal{I}}}

\newcommand*{\inputQuote}[1]{\ensuremath{\mathcal{Q}_{#1}}}

\newcommand*{\sharedGreaterThan}{\ensuremath{\stackrel{?}{>}}}

\newcommand*{\sharedMult}{\ensuremath{ \ \cdot \ }}
\newcommand*{\sharedSubtract}{\ensuremath{-}}
\newcommand*{\sharedAdd}{\ensuremath{+}}
\newcommand*{\sharedScalMult}{\ensuremath{\times}}
\newcommand*{\compCheck}{\ensuremath{\small{\textsc{Comp-Check}}}}

\newcommand*{\simplexGate}{\ensuremath{\small{\textsc{Simplex}}\small}}
\newcommand*{\ipUpdateGate}{\ensuremath{\small{\textsc{IP-Update}}\small}}
\newcommand*{\ipSetupGate}{\ensuremath{\small{\textsc{IP-Setup}}\small}}
\newcommand*{\ipBranchGate}{\ensuremath{\small{\textsc{IP-Branch}}\small}}
\newcommand*{\branchAndBoundProtocol}{\ensuremath{\small{\textsc{Branch-and-Bound}}}}
\newcommand*{\minSelGate}{\ensuremath{\small{\textsc{Sel-Min}}\small}}
\newcommand*{\kepIpProtocol}{\ensuremath{\small{\textsc{Kep-IP}}}}
\newcommand*{\kepIpFunc}{\ensuremath{\func_{\small{\textsc{Kep-IP}}}}}
\newcommand*{\shuffleGate}{\ensuremath{\small{\textsc{Shuffle}}}}
\newcommand*{\reverseShuffleGate}{\ensuremath{\small{\textsc{Reverse-Shuffle}}}}

\newcommand*{\bitsize}{\ensuremath{s}}
	
\newcommand*{\donorbloodvec}[1]{\ensuremath{B^{d}_{#1}}}
\newcommand*{\donorbloodvecNotSpec}{\ensuremath{B^{d}}}
\newcommand*{\patientbloodvec}[1]{\ensuremath{B^{p}_{#1}}}

\newcommand*{\antigenvec}[1]{\ensuremath{A^{d}_{#1}}}

\newcommand*{\antibodyvec}[1]{\ensuremath{A^{p}_{#1}}}
\newcommand*{\antibodyvecNotSpec}{\ensuremath{A^{p}}}

\definecolor{smartgray}{RGB}{61, 62, 64}
\definecolor{smartblue}{HTML}{0B6296}
\definecolor{smartred}{HTML}{C54949}
\definecolor{smartyellow}{HTML}{d0d62a}
\definecolor{smartgreen}{HTML}{61c15d}
\definecolor{smartorange}{HTML}{ff7e00}
\definecolor{smartviolet}{HTML}{d19fe8}
\definecolor{smartbrown}{HTML}{480607}
\definecolor{smartcadet}{HTML}{5F9EA0}

\newcommand*{\ints}{\ensuremath{\mathbb{Z}}}
\newcommand*{\nats}{\ensuremath{\mathbb{N}}}

\newcommand*{\reals}{\ensuremath{\mathbb{R}}}

\newcommand*{\draw}{\ensuremath{\leftarrow}}

\newcommand*{\bigo}[1]{\ensuremath{\mathcal{O}(#1)}}

\newcommand*{\enc}[1]{\ensuremath{[ #1 ]}}

\newcommand*{\partysymbol}[1]{\ensuremath{P_{#1}}}
\newcommand*{\parties}{\ensuremath{\mathcal{P}}}

\newcommand*{\numparties}{\ensuremath{\iota}}

\newcommand*{\func}{\ensuremath{\pazocal{F}}}

\newcommand*{\simval}[1]{\ensuremath{\langle #1 \rangle}}


%% file: settings.tex
\newcommand\Tstrut{\rule{0pt}{2.2ex}}         
\newcommand\TstrutLarge{\rule{0pt}{3ex}}         
\newcommand\Bstrut{\rule[-0.9ex]{0pt}{0pt}}   
\newcommand\BstrutLarge{\rule[-1.5ex]{0pt}{0pt}}   

\newtheorem{thm}{Theorem}
\newtheorem{definition}[thm]{Definition}

\newtheorem{functionality}{Functionality}

\makeatletter
\newcounter{protocol}
\newenvironment{protocol}[1][]{
	\let\c@algorithm\c@protocol
	\makeatletter
	\renewcommand{\ALG@name}{Protocol}
	\makeatother
	\begin{algorithm}[#1]
	}{\end{algorithm}
}
\makeatother



\makeatletter
\newcounter{gate}

\makeatother

\makeatletter
\newcounter{alg}

\newcolumntype{L}[1]{>{\raggedright\let\newline\\\arraybackslash\hspace{0pt}}m{#1}}
\newcolumntype{C}[1]{>{\centering\let\newline\\\arraybackslash\hspace{0pt}}m{#1}}
\newcolumntype{R}[1]{>{\raggedleft\let\newline\\\arraybackslash\hspace{0pt}}m{#1}}

\tikzset{every tree node/.style={align=center}}

%% file: chapters/abstract.tex
The kidney exchange problem (KEP) is to find a constellation of exchanges that maximizes the number of transplants that can be carried out for a set of pairs of patients with kidney disease and their incompatible donors. Recently, this problem has been tackled from a privacy perspective in order to protect the sensitive medical data of patients and donors and to decrease the potential for manipulation of the computing of the exchanges. 
However, the proposed approaches to date either only compute an approximative solution to the KEP or they suffer from a huge decrease in performance.
In this paper, we suggest a novel privacy-preserving protocol that computes an exact solution to the KEP and significantly outperforms the other existing exact approaches. 
Our novel protocol is based on Integer Programming which is the most efficient method for solving the KEP in the non privacy-preserving case. 
We achieve an improved performance compared to the privacy-preserving approaches known to date by extending the output of the ideal functionality to include the termination decisions of the underlying algorithm.
We implement our protocol in the SMPC benchmarking framework MP-SPDZ and compare its performance to the existing protocols for solving the KEP. 
In this extended version of our paper, we also evaluate whether and if so how much information can be inferred from the extended output of the ideal functionality.

%% file: chapters/introduction.tex
\section{Introduction}

In living kidney donation, a patient tries to find a friend or relative who is willing to donate one of her kidneys to the patient. Even if a patient finds such a living donor, this donor is often not medically compatible with the patient. Kidney exchange tries to solve this problem by finding constellations in which multiple such incompatible \emph{patient-donor pairs} can exchange their kidney donors among each other. These exchanges are typically carried out in \emph{exchange cycles} where the donor of a patient-donor pair always donates to the patient of the succeeding pair in the cycle and the patient of a pair receives a kidney donation from the donor of the preceding pair. In practice, the maximum size of these cycles is commonly restricted to two or three due to the amount of medical resources required to carry out a transplant~\cite{AshlagiKidneyExchangeOperations2021}.

More formally, the \emph{Kidney Exchange Problem} (KEP) is to find an \emph{optimal} set of exchange cycles \emph{up to} a maximum cycle size $ \cycleBound $ for a fixed set of patient-donor pairs such that the number of possible transplants that can be carried out is maximized~\cite{AbrahamClearingAlgorithms2007}.

In many countries, there are already central platforms that facilitate the solving of the KEP for registered incompatible patient-donor pairs~\cite{Biro_EuropeanKE_2019,AshlagiKidneyExchangeOperations2021}. The drawbacks of this centralized approach are that these platforms possess the medical data of all registered patient-donor pairs and that they exhibit complete control over the computation of the possible exchanges. Thus, a compromise of such a central platform would have severe consequences for the registered patient-donor pairs.

In recent research, initial approaches to address these shortcomings of the existing platforms have been proposed~\cite{Breuer_KEprotocol_2020,Breuer_Matching_2022}. These approaches use \emph{Secure Multi-Party Computation} (SMPC) to solve the KEP in a distributed and privacy-preserving fashion. In particular, they implement the ideal functionality that obtains the private medical data of the patient-donor pairs as input and for each patient-donor pair outputs its possible exchange partners such that the overall number of possible transplants is maximized. Still, these initial approaches exhibit some drawbacks. While the protocol from~\cite{Breuer_Matching_2022} only solves the KEP for a maximum cycle size $ \cycleBound = 2 $, the protocol from~\cite{Breuer_KEprotocol_2020} allows for an arbitrary maximum cycle size but only yields practical runtimes for a very small number of incompatible patient-donor pairs.\footnote{The KEP is NP-complete for $ \cycleBound \geq 3 $~\cite{AbrahamClearingAlgorithms2007}.}

In this paper, we aim to address these drawbacks.
To this end, we turn to the most efficient algorithms employed in the non-privacy-preserving setting to solve the KEP, which are based on Integer Programming (IP)~\cite{AbrahamClearingAlgorithms2007}. Even though these algorithms have exponential time complexity in the worst case~\cite{Schrijver_LinearProgramming_1998}, they have been shown to efficiently solve the KEP in practice~\cite{AbrahamClearingAlgorithms2007}. Therefore, we explore whether and if so how they can be used to devise an efficient privacy-preserving protocol for solving the KEP for an arbitrary maximum cycle size $ \cycleBound $.

\textbf{Approach.} As the basis of our novel privacy-preserving protocol, we use the Branch-and-Bound algorithm introduced in~\cite{Land_BranchAndBound_1960}. At its core, this algorithm builds a tree of subproblems whose branches are pruned as soon as the upper bound on solutions in a branch is worse than the best known solution so far. This makes it more efficient than a brute force approach that explores the complete search space. Each subproblem in the tree requires the solving of a Linear Program (LP). In our protocol, we solve these LPs using an existing privacy-preserving implementation of the Simplex algorithm~\cite{Toft_PPLinearProgramming_2009},
which is the most prominent algorithm for solving LPs~\cite{Schrijver_LinearProgramming_1998}.

It is hard to predict the number of iterations that IP- and LP-based approaches require to find an optimal solution. Therefore, the output of the ideal functionality of the respective SMPC protocols typically includes additional information. For instance, the output of the ideal functionality of the SMPC protocols known to date for the Simplex algorithm include the number of iterations that Simplex requires to find an optimal solution~\cite{Catrina_PPLinearProgramming_2010,Toft_PPLinearProgramming_2009}. Since we use the protocol from~\cite{Toft_PPLinearProgramming_2009} as a building block, the output of the ideal functionality for our novel privacy-preserving protocol includes the number of Simplex iterations for each of the subproblems in the Branch-and-Bound tree. Furthermore, we deliberately include the structure of the Branch-and-Bound tree corresponding to the pruning decisions of the Branch-and-Bound algorithm in the output of the ideal functionality in order to increase the runtime performance of our novel protocol.\footnote{Note that these extended outputs do not compromise the security of our novel protocol~(cf.~Section~\ref{sub:kep_protocol}).} 

\textbf{Contributions.}
Our paper provides the following three main contributions: 
First, we devise an efficient SMPC protocol for solving the KEP for an arbitrary maximum cycle size $ \cycleBound $ based on the Branch-and-Bound algorithm and prove its security (Section~\ref{sec:protocol}).
Second, we implement our novel protocol in the SMPC benchmarking framework MP-SPDZ~\cite{KellerMPSPDZ2020} and evaluate its performance based on a real-world data set from a large kidney exchange platform in the USA (Section~\ref{sec:evaluation}).
Third, we compare the implementation of our protocol to the protocol from~\cite{Breuer_KEprotocol_2020} which to date is the only known SMPC protocol for solving the KEP for an arbitrary maximum cycle size~$ \cycleBound $~\cite{Breuer_KEprotocol_2020}. We show that our protocol allows for a significant improvement of the runtime (Section~\ref{sec:evaluation}).

\textbf{Extended Version.}
Compared to the original version of this paper~\cite{Breuer_KepIp_2022}, we also evaluate whether the deliberate inclusion of the structure of the Branch-and-Bound tree and the number of Simplex iterations in the output of the ideal functionality allows to infer any further information compared to the output of the existing protocols for solving the KEP~\cite{Breuer_KEprotocol_2020,Breuer_Matching_2022} (cf.~Appendix~I).
Based on our findings, we present techniques for limiting the information that can be inferred from the structure of the Branch-and-Bound tree and the number of Simplex iterations (cf.~Appendix~II).
Finally, we include an updated runtime evaluation of our protocol. Using an increased degree of parallelization in MP-SPDZ, replacing Shamir secret sharing~\cite{ShamirSecretSharing1979} by the replicated secret sharing scheme from~\cite{Araki_ReplicatedSecretSharing_2016}, and updating the MP-SPDZ version from 0.2.5 to 0.3.5 (cf.~Section~\ref{sec:evaluation}) lead to a significant improvement of the runtime performance compared to the evaluation presented in the original version of this paper.

%% file: chapters/preliminaries.tex
\section{Preliminaries}

In this section, we provide the formal background on which our privacy-preserving protocol for solving the Kidney Exchange Problem (KEP) is based. In Section~\ref{sub:notation}, we introduce the notation used in this paper. In Section~\ref{sub:smpc}, we describe the Secure Multi-Party Computation (SMPC) setting for our protocols. In Section~\ref{sub:integer_programming}, we give an introduction to Integer Programming (IP) in the context of kidney exchange. 

\subsection{Notation}\label{sub:notation}
A \emph{directed graph} is a graph $ G = (\nodes, E) $ where the edge $ (v, w) \in E $ ($ v, w \in \nodes $) is directed from node~$ v $ to node~$ w $. Such a graph is usually represented as an \emph{adjacency matrix}~$ \compGraphMatrix $ of size $ \vert \nodes \vert \times \vert \nodes \vert $ such that $ M(i, j) = 1 $ if $ (i, j) \in E $ and $ M(i, j) = 0 $, otherwise. Given a matrix $ M $, we write $ M(i, -) $ for its $ i $-th row, $ M(-, j) $ for its $ j $-th column, and $ M([i_1 : i_2], [j_1 : j_2]) $ for the submatrix with entries $ M(i, j) $ where $ i_1 \leq i \leq i_2 $ and $ j_1 \leq j \leq j_2 $.

We adopt the notation for kidney exchange from~\cite{Breuer_KEprotocol_2020}. In particular, we denote the index set of all patient-donor pairs $ \partysymbol{1}, ..., \partysymbol{\numparties} $ ($ \numparties \in \nats $) by $ \parties := \{1, ..., \numparties\} $. We call the graph induced by the patient-donor pairs' medical input the \emph{compatibility graph} $ G = (V, E) $ where $ V = \parties $ and $ (u, v) \in E $ if the donor of pair $ \partysymbol{u} $ can donate to the patient of pair $ \partysymbol{v} $. An \emph{exchange cycle} of size $ m $ is a tuple $ \singleCycle = (\partysymbol{i_1}, ..., \partysymbol{i_m}) $ with $ i_j \neq i_k $ such that for $ j \in \{1, ..., m-1\} $ it holds that $ (v_j, v_{j+1}) \in E $ and $ (v_m, v_1) \in E $.

\subsection{Secure Multi-Party Computation}\label{sub:smpc}

In SMPC a set of $ \numCompPeers $ parties computes a functionality in a distributed way such that each party does not learn anything beyond its private input and output and what can be deduced from both. We adopt the setting for privacy-preserving kidney exchange from Breuer et al.~\cite{Breuer_Matching_2022}. In particular, we distinguish between \emph{computing peers} and \emph{input peers}. The computing peers execute the actual SMPC protocol for computing the functionality among each other, whereas the input peers only provide private input to the functionality and receive their private output. In the kidney exchange setting, each patient-donor pair forms one input peer and the computing peers can be run by large transplant centers or governmental organizations.

Our SMPC protocols require any $ (t,\numCompPeers) $ threshold linear secret sharing scheme over $ \ints_M $, where $ M $ is a large prime $ p $, or $ 2^k $ for a large integer $ k $. The threshold $ t $ indicates that at least $ t+1 $ shares are required to reconstruct the shared value. We keep our protocol independent from any particular scheme by considering an \emph{arithmetic black box} (ABB)~\cite{Damgard_ABB_2003}. The ABB enables the sharing and reconstruction of secret values, and the computation of linear combinations as well as multiplications of secret values. While the computing peers can compute linear combinations of shared values locally, their multiplication requires communication between the computing peers.
We denote a shared value $ x $ by $ \enc{x} $, a vector $ V $ of shared values by $ \enc{V} $, and a matrix $ M $ of shared values by~$ \enc{M} $. We write $ \enc{V}(i) $ (resp.~$ \enc{M}(i,j) $) to denote a single element of a secret vector $ \enc{V} $ (resp.~matrix $ \enc{M} $). 
We denote the multiplication of two shared values $ \enc{x} $ and $ \enc{y} $ by $ \enc{z} \leftarrow \enc{x} \cdot \enc{y} $. 

\customParagraph{Complexity metrics.} 
We use the two common complexity metrics communication and round complexity. The communication complexity is determined by the amount of data that is sent during a protocol execution. The round complexity corresponds to the number of sequential communication steps required by a protocol. Note that steps that can be executed in parallel are counted as a single round.

\begin{protocol}[t]
	\caption{$ \minSelGate $}
	\label{prot:min_sel}
	\input{./protocols/min_sel.tex}

\end{protocol}
\customParagraph{Building Blocks.}
For our protocol~$ \kepIpProtocol $ (cf.~Section~\ref{sec:protocol}), we require privacy-preserving protocols for several basic primitives. For the secure comparison of two shared values $ \enc{x} $ and $ \enc{y} $ we write $ \enc{x} \sharedGreaterThan \enc{y} $. A protocol for this primitive can be realized with linear communication and logarithmic round complexity~\cite{Catrina_PrimitivesSMPC_2010,Escudero_edaBits_2020}. Conditional selection is defined as choosing between two values based on a secret shared bit~$ \enc{b} $, i.e., we choose $ \enc{x} $ if $ b $ is equal to $ 1 $ and $ \enc{y} $, otherwise. This corresponds to computing $ \enc{b} \cdot (\enc{x} - \enc{y}) + \enc{y} $ and we denote this by $ \enc{b} \ ? \ \enc{x} : \enc{y} $. Finally, we require a protocol $ \minSelGate $ for selecting the index of the first element in a vector that is larger than zero. The specification of this protocol is given in Protocol~\ref{prot:min_sel}. The communication complexity of $ \minSelGate $ is $ \bigo{\vert U \vert \bitsize} $ and its round complexity is $ \bigo{\log(\bitsize)} $, where $ \bitsize $ is the bit length of the input to the comparison protocol.

\customParagraph{Security Model.}
Same as Breuer et al.~\cite{Breuer_Matching_2022} we prove our protocol to be secure in the \textit{semi-honest model}. This means that we assume a set of corrupted computing peers who strictly follow the protocol specification but try to deduce as much information on the honest computing peers' input as possible. As security in the semi-honest model already prevents an adversary from learning the medical data of the patient-donor pairs and thereby from influencing the computed exchanges in any meaningful way, we consider it to be sufficient for our use case. Besides, the input peers can not learn anything beyond their private input and output since they are not part of the protocol execution itself. Finally, we consider an honest majority of computing peers and the existence of encrypted and authenticated channels between all participating parties. For the security proofs of our protocols, we use the standard simulation-based security paradigm. We denote the values simulated by the simulator by angle brackets $ \simval{\cdot} $. For further details on the semi-honest security model and the simulation-based security paradigm, we refer the reader to~\cite{GoldreichFoundationsTwo2004}.

\subsection{Integer Programming for the Kidney Exchange Problem}\label{sub:integer_programming}

A \emph{Linear Program (LP)} is a formulation for an optimization problem of the following form: Let $ c \in \reals^n, a_i \in \reals^n $, and $ b_i \in \reals $ for $ i \in \{1, ..., \numColumns\} $. The LP is then defined as 
\begin{align}\label{eq:lp}
	\begin{split}
		\text{maximize }	& \objectiveVec^\top x				\\
		\text{s.t. }		& \constraintMatrix x = \boundVec	\\
							& x \ge 0,
	\end{split}
\end{align}
where $ \objectiveFunc $ is the \emph{objective function}, $ \constraintMatrix $ is the \emph{constraint matrix}, and $ \boundVec $ is the \emph{bound vector}. This form of an LP where only equality constraints occur is called the \emph{standard form}. Inequalities $ \constraintEntry_i^\top x \leq \boundVec_i $ can be represented in standard form as $ \constraintEntry_i^\top x + \slackVector_i = \boundVec_i $ where $ s_i \geq 0 $ is called a \emph{slack variable}.

In our protocols, we represent an LP in form of a \emph{tableau}
\begin{align}\label{eq:tableau-structure}
	T=\left(\begin{array}{ccc|c}
		&c^\top	& 	& z	\\\hline 
		& A		& 	& b \TstrutLarge\BstrutLarge
	\end{array}\right)
\end{align}
with objective value $ -\objectiveValue = \objectiveVec^\top \solutionVec $ (the sign is flipped as the tableau method is designed for minimization whereas we aim at maximizing the objective function).

In contrast to an LP, for IP the additional restriction that all solutions have to be integers (i.e., $ x \in \nats^n $) is imposed. This is a natural requirement for the KEP as an exchange cycle is either part of the solution or not.

\customParagraph{Cycle Formulation.}
For the restricted KEP where only cycles of bounded length $ \cycleBound $ are allowed, the most efficient known IP formulation is the cycle formulation where a constraint is introduced for each possible cycle of length at most~$ \cycleBound $~\cite{AbrahamClearingAlgorithms2007}. 

\begin{definition}[Cycle Formulation]
	\textit{
		Let $ \cycles $ be the set of all cycles of length at most $ \cycleBound $ in the compatibility graph $ G = (V, E) $. Then, the \emph{cycle formulation} is defined as
		\begin{equation}\label{eq:cycle_formulation}
			\begin{alignedat}{3}
				\text{maximize }& \sum_{\singleCycle \in \cycles} \vert \singleCycle \vert \ \cycleVariable														\\
				\text{s.t. }	& \sum_{\substack{\singleCycle \in \cycles \\ v \in V(\singleCycle)}} \cycleVariable \leq 1	&& \text{ for all }v \in V		\\
								& \cycleVariable \in \{0,1\}											&& \text{ for all }\singleCycle \in \cycles
			\end{alignedat}
		\end{equation}
		where $ \cycleVariable \in \{0, 1\} $ for $ \singleCycle \in \cycles $ such that $ \cycleVariable = 1 $ iff cycle $ \singleCycle $ is chosen and $ \cycleVariable = 0 $, otherwise. 
	}
\end{definition}

\customParagraph{Subset Formulation.}
In the privacy-preserving setting, we have to keep secret which cycles are present in the compatibility graph. Therefore, we have to consider the set $ \cycles $ of all \emph{potential} cycles in the complete directed graph $ K = (V, E') $ 
with $ V = \parties $ and $ (u, v) \in E' $ for all $ u, v \in V $. Thus, if two cycles $ \singleCycle_1, \singleCycle_2 \in \cycles $ contain the same vertices in a different order, their coefficients for the objective function in the cycle formulation are exactly the same. We eliminate this redundancy by proposing the more efficient \emph{subset formulation}.\footnote{This formulation is only more efficient in the privacy-preserving case as only there the complete graph~$ \completeGraph $ has to be considered.}

\begin{definition}[Subset Formulation]
	\textit{
		Let $ \subsets $ contain all vertex sets of size $ k $ with $ 2 \leq k \leq \cycleBound $ and let $ \subsetMap: \subsets \rightarrow \cycles \cup \{\emptyset\} $ be a mapping that assigns an arbitrary cycle $ \subsetMap(\singleSubset) \in \cycles $ with $ \nodes(\subsetMap(\singleSubset)) = \singleSubset $ to each $ \singleSubset \in \subsets $. If there is no such cycle, let $ \subsetMap(\singleSubset) = \emptyset $. Then, the \emph{subset formulation} is defined as
		\begin{equation}\label{eq:subset_formulation}
			\begin{alignedat}{3}
				\text{maximize }	& \sum_{\singleSubset \in \subsets} \vert \subsetMap(\singleSubset) \vert \ \subsetVariable	\\
				\text{s.t. }		& \sum_{\substack{\singleSubset \in \subsets \\ v \in \singleSubset}} \subsetVariable \leq 1 && \text{ for all } v \in \nodes \\
									& \subsetVariable \in \{0, 1\} && \text{ for all } \singleSubset \in \subsets
			\end{alignedat}
		\end{equation}
		where the variable $ \subsetVariable \in \{0, 1\} $ indicates for each $ \singleSubset \in \subsets $ whether any cycle consisting of the vertices in $ \singleSubset $ is chosen.
	}
\end{definition}

For our protocol, we assume that $ \subsets = \{\singleSubset_1, ..., \singleSubset_{\vert \subsets \vert}\} $ is a fixed enumeration of the subsets and for $ 1 \leq i \leq \vert \subsets \vert $ we have an enumeration $ \cycles_i = \{\singleCycle_{i, 1}, ..., \singleCycle_{i, \completeGraphSize}\} $ of the cycles in the complete graph~$ \completeGraph $.

Observe that the number of subsets is smaller by a factor of $ (\cycleBound - 1)! $ as there are $ (\cycleBound - 1)! $ cycles on a set of size $ \cycleBound $. Hence, we have to consider significantly less variables at the initial cost of determining the mapping~$ \subsetMap $ from subsets to cycles.

\customParagraph{Example.}
For an intuition on the relation between the initial tableau of the LP relaxation\footnote{The LP relaxation of an IP is the LP that we obtain if we remove the integrality constraint on the variables of the IP.} for the subset formulation and the compatibility graph, consider the example compatibility graph provided in Figure~\ref{fig:subset_formulation_example}. To build the initial tableau for the subset formulation for this graph, we add one column for all subsets of nodes that could form a cycle between four nodes. For our example, this yields the subset variables $ \explicitSubsetVar{12}, \explicitSubsetVar{13}, \explicitSubsetVar{14}, \explicitSubsetVar{23}, \explicitSubsetVar{24}, \explicitSubsetVar{34}, \explicitSubsetVar{123}, \explicitSubsetVar{124}, \explicitSubsetVar{134}, \explicitSubsetVar{234} $ and the initial tableau $ T := $
\begin{align*}
	 \left(\begin{array}{rrrrrrrrrrrrrr|r}
		0 &	0 &	0 &	2	& 0	& 0	& 0	& 3	& 0	& 0	& 0	& 0	& 0	& 0	& 0	\\\hline
		1 &	1 &	1 &	0	& 0	& 0	& 1	& 1	& 1	& 0	& 1	& 0	& 0	& 0	& 1	\Tstrut\\
		1 &	0 &	0 &	1	& 1	& 0	& 1	& 1	& 0	& 1	& 0	& 1	& 0	& 0 & 1	\\
		0 &	1 &	0 &	1	& 0	& 1	& 1	& 0	& 1	& 1	& 0	& 0	& 1	& 0	& 1	\\
		0 &	0 &	1 &	0	& 1	& 1	& 0	& 1	& 1	& 1	& 0	& 0	& 0	& 1	& 1	
	\end{array}\right)
\end{align*}

\begin{figure}[t]
	\input{./figures/subset_formulation_example.tex}
	\caption{Example of a compatibility graph and its exchange cycles.}
	\label{fig:subset_formulation_example}
\end{figure}

%% file: protocols/min_sel.tex
\algrenewcommand\algorithmicindent{1em}
\small
\textbf{Input:} A vector $ \enc{U} $ with $ U(i) \in \{0, 1\} $. \vspace*{0.2em} \\
\textbf{Output:} A vector $ \enc{W} $ with $ W(i) = 1 $ if $ U(1) = ... = U(i-1) = 0 $ and $ U(i) = 1 $, and $ W(i) = 0 $ otherwise.
\begin{algorithmic}[1]
	\For{$ i = 1, ..., \vert U \vert $}
		\State $ \enc{W}(i) \leftarrow \enc{U}(i) \sharedGreaterThan \sum_{j = 1}^{i-1}\enc{U}(j) $
	\EndFor
\end{algorithmic}

%% file: figures/subset_formulation_example.tex
\centering
\begin{tikzpicture}[ 
	> = stealth, 
	shorten > = 1pt, 
	shorten < = 1pt,
	auto,
	node distance = 2cm, 
	semithick 
	]
	\def\distance{1.2}
	\def\graphdistance{3}
	\def\labelsize{\scriptsize}
	\def\vertdistance{2.4}
	
	\node[anchor=center] (compgraph) at (0.5*\distance, 0.6) {\small\underline{Compatibility Graph}};
	
	\node[draw, circle, minimum size=0.25cm, inner sep=1pt] (p1) at (0, 0) {\labelsize$ \partysymbol{1} $};
	\node[draw, circle, minimum size=0.25cm, inner sep=1pt] (p2) at (\distance, 0) {\labelsize$ \partysymbol{2} $};
	\node[draw, circle, minimum size=0.25cm, inner sep=1pt] (p3) at (0, -\distance) {\labelsize$ \partysymbol{3} $};
	\node[draw, circle, minimum size=0.25cm, inner sep=1pt] (p4) at (\distance, -\distance) {\labelsize$ \partysymbol{4} $};
	
	\path[->] (p1) edge (p2);
	\path[->, bend left=15] (p2) edge (p3);
	\path[->] (p2) edge (p4);
	\path[->, bend left=15] (p3) edge (p2);
	\path[->] (p4) edge (p1);
	\path[->] (p1) edge (p2);
	
	\node[anchor=center] (compgraph) at (1.5*\graphdistance, 0.6) {\small\underline{Exchange Cycles}};
	
	\node[draw, circle, minimum size=0.25cm, inner sep=1pt] (c1p1) at (\graphdistance+0, 0) {\labelsize$ \partysymbol{1} $};
	\node[draw, circle, minimum size=0.25cm, inner sep=1pt] (c1p2) at (\graphdistance+\distance, 0) {\labelsize$ \partysymbol{2} $};
	\node[draw, circle, minimum size=0.25cm, inner sep=1pt] (c1p3) at (\graphdistance+0, -\distance) {\labelsize$ \partysymbol{3} $};
	\node[draw, circle, minimum size=0.25cm, inner sep=1pt] (c1p4) at (\graphdistance+\distance, -\distance) {\labelsize$ \partysymbol{4} $};
	
	\path[->, bend left=15] (c1p2) edge (c1p3);
	\path[->, bend left=15] (c1p3) edge (c1p2);

	\node[draw, circle, minimum size=0.25cm, inner sep=1pt] (c2p1) at (1.7*\graphdistance, 0) {\labelsize$ \partysymbol{1} $};
	\node[draw, circle, minimum size=0.25cm, inner sep=1pt] (c2p2) at (1.7*\graphdistance+\distance, 0) {\labelsize$ \partysymbol{2} $};
	\node[draw, circle, minimum size=0.25cm, inner sep=1pt] (c2p3) at (1.7*\graphdistance, -\distance) {\labelsize$ \partysymbol{3} $};
	\node[draw, circle, minimum size=0.25cm, inner sep=1pt] (c2p4) at (1.7*\graphdistance+\distance, -\distance) {\labelsize$ \partysymbol{4} $};
	
	\path[->] (c2p1) edge (c2p2);
	\path[->] (c2p2) edge (c2p4);
	\path[->] (c2p4) edge (c2p1);

\end{tikzpicture}

%% file: chapters/branch_and_bound.tex
\section{Branch-and-Bound Algorithm}\label{sec:branch_and_bound}

Our protocol $ \kepIpProtocol $ (cf.~Section~\ref{sub:kep_protocol}) for solving the KEP is based on the Branch-and-Bound algorithm~\cite{Land_BranchAndBound_1960} which is a common algorithm for IP. 
The main problem that is solved by the Branch-and-Bound algorithm is that we may obtain fractional solutions if we just solve the LP relaxation of the subset formulation using an LP solver such as the Simplex algorithm~\cite{Dantzig_LinearProgrammingSimplex_1963}. For the KEP, however, it has to hold that all solutions are integral as a cycle can either be part of the solution or not, i.e., a variable can either have value 1 or 0. 

The core idea of the Branch-and-Bound algorithm is to repeatedly run an LP solver until an integral solution of optimal value is found. 
More specifically, the Branch-and-Bound algorithm generates a tree of subproblems where each node of the tree states an upper bound on a possible optimal solution. Each node also stores the best known integer solution $ \incumbent \in \ints^n $ (called the \emph{incumbent}) and its value $ \incumbentValue \in \reals $. The solution of a subproblem is then determined using an LP solver. Throughout this paper, we use the Simplex algorithm as it is the most prominent method for solving LPs~\cite{Schrijver_LinearProgramming_1998} and there are already SMPC protocols for it (cf.~Section~\ref{sub:rw_pp_lp}). 

\begin{algorithm}[t]
	\caption{Branch-and-Bound algorithm, based on~\cite{Land_BranchAndBound_1960}.}
	\label{alg:branch_and_bound}
	\input{./protocols/plain_branch_and_bound.tex}
\end{algorithm}

Algorithm~\ref{alg:branch_and_bound} contains pseudocode for the Branch-and-Bound algorithm. At the beginning, an initial tableau $ \initialTableau $ of the LP relaxation for the IP is built. Together with the initial upper bound $ \upperbound = \infty $, this tableau forms the first subproblem in the list of subproblems $ L $ (i.e., it forms the root node of the Branch-and-Bound tree). The algorithm then solves the subproblems in the list $ L $ one by one until $ L $ is empty. For each subproblem $ (\tableau, \upperbound) $, it is first checked whether the best known solution $ \incumbentValue $ is larger than the upper bound $ \upperbound $ of the subproblem. If this is the case, this subproblem cannot improve the solution and the whole subtree is pruned. Otherwise, the LP with tableau $ T $ is solved yielding a solution $ \simplexSolutionComplete $ of value $ \simplexSolutionValue $. If the solution is integral, the incumbent $ \incumbent $ is updated if this improves its value (i.e., if $ \simplexSolutionValue > \incumbentValue $). If the solution is fractional, one currently fractional variable $ x_i $ is chosen and two new child nodes $ (\tableau_1, \simplexSolutionValue) $ and $ (\tableau_2, \simplexSolutionValue) $ are created such that the new value of $ x_i $ is set to $ \lfloor \simplexSolutionComplete(i) \rfloor $ for one child node and to $ \lceil \simplexSolutionComplete(i) \rceil $ for the other. Both new subproblems are added to the list $ L $. The algorithm terminates when $ L $ is empty.

Figure~\ref{fig:bb_tree} shows a Branch-and-Bound tree for the IP
\begin{equation}\label{eq-example-ip-bb}
	\begin{alignedat}{2}
		\max\:		& z = 3 x_1 + 2 x_2 + 2 x_3	\\
		\text{s.t. }& x_1 + 2 x_2 + 3 x_3 \le 3	\\
					& 3 x_1 + x_2 + 2 x_3 \le 3	\\
					& 2 x_1 + 3 x_2 + x_3 \le 3	\\
					& x_1, x_2, x_3 \ge 0.
	\end{alignedat}
\end{equation}
Note that the subproblem at Node~2 has an integer solution. Since the upper bound $ \upperbound $ of the subproblems at Nodes 4 and~5 is lower than the incumbent value $ \incumbentValue $, they do not have to be evaluated, i.e., their whole subtrees can be pruned.

\begin{figure}[t]
	\input{figures/branch_and_bound_tree.tex}
	\vspace*{-2em}
	\caption{Example Branch-and-Bound tree for the IP in Equation~\ref{eq-example-ip-bb}. The order~in which subproblems are evaluated and their bounds are given above the nodes.}
	\label{fig:bb_tree}
\end{figure}

\customParagraph{Simplifications for the KEP.}
The subset formulation (Equation~\ref{eq:subset_formulation}, Section~\ref{sub:integer_programming}) allows us to make several simplifying assumptions when using Algorithm~\ref{alg:branch_and_bound} for solving the KEP. We can use the feasible initial solution $ x =  (0,\ldots,0)^\top $ of value $ 0 $ as the starting point $ (\incumbent, \incumbentValue) $. This corresponds to selecting no exchange cycle at all. Furthermore, we can use the initial upper bound $ \numparties $ (instead of $ \infty $) for the initial subproblem as the best solution for the KEP is that all patient-donor pairs are included in the computed exchange cycles. Finally, the LP relaxation of the subset formulation already guarantees that $ 0 \le x_i \le 1$ for all variables $ x_i $. Thus, the new constraints in Step 6 of Algorithm~\ref{alg:branch_and_bound} are always $ x_i = 0 $ and $ x_i = 1 $. Note that adding the constraints $ x_i = 0 $ or $ x_i = 1 $ cannot lead to an infeasible problem as we can always obtain a feasible point where $ x_i = 1 $ by setting all other fractional variables to $ 0 $. 

%% file: protocols/plain_branch_and_bound.tex
\algrenewcommand\algorithmicindent{1em}
\small
\begin{algorithmic}[1]
	\State Let $ L = \{(\tableau, \infty)\} $ where $ \tableau $ is the initial tableau of the LP relaxation with $ \incumbent = \bot $, and $ \incumbentValue = -\infty $.
	\vspace*{0.1em}
	\State If $ L $ is empty, return $ \incumbent $. Otherwise, choose and remove a problem $ (T, \upperbound) $ from $ L $.
	\vspace*{0.1em}
	\State If $ \incumbentValue \geq \upperbound $, go to Step 2.
	\vspace*{0.1em}
	\State Solve the LP with tableau $ \tableau $ obtaining a solution $ \simplexSolutionComplete \in \reals^n $ with value $ \simplexSolutionValue \leq \upperbound $.
	\vspace*{0.1em}
	\State If $ \simplexSolutionComplete \in \ints^n $ and $ \simplexSolutionValue > \incumbentValue $, replace $ \incumbent $ with $ \simplexSolutionComplete $ and $ \incumbentValue $ with $ \simplexSolutionValue $.
	\vspace*{0.1em}
	\State If $ \simplexSolutionComplete \notin \nats^n $, choose an index $ i $ with $ \simplexSolutionComplete(i) \notin \nats $, create new tableaux $ T_1 $ and $ T_2 $ from $ T $ by adding the constraint $ x_i \leq \lfloor \simplexSolutionComplete(i) \rfloor $ for $ \tableau_1 $ and $ x_i \geq \lceil \simplexSolutionComplete(i) \rceil $ for $ \tableau_2 $, and set $ L \leftarrow L \cup \{(T_1, \simplexSolutionValue), (T_2, \simplexSolutionValue)\} $.
	\vspace*{0.1em}
	\State Go to Step $ 2 $.
\end{algorithmic}

%% file: figures/branch_and_bound_tree.tex
\begin{adjustbox}{width=\columnwidth}
	\begin{tikzpicture}
		\def\vertdist{1.8}
		\def\horidist{3}
		\node[align=center] (a) at (0, 0) {1. ($ \upperbound = \infty, \incumbentValue = 0 $)\\\fbox{$ \simplexSolutionComplete = (\frac12, \frac12, \frac12), \simplexSolutionValue = \frac72 $}};
		\node[align=center] (b) at (-\horidist, -\vertdist) {2. ($ \upperbound = \frac72, \incumbentValue = 0 $)\\\fbox{$ \simplexSolutionComplete = (1,0,0), \simplexSolutionValue = 3 $}};
		\node[align=center] (c) at (\horidist, -\vertdist) {3. ($ \upperbound = \frac72, \incumbentValue = 3 $)\\\fbox{$ \simplexSolutionComplete = (0, \frac67, \frac37), \simplexSolutionValue = \frac{18}{7}$}};
		\draw (a) edge node[midway,left] {$x_1\ge1 \ $} (b);
		\draw (a) edge node[midway,right] {$ \ x_1\le0$} (c);
		\node[align=center] (d) at (0, -2*\vertdist) {4. ($ \upperbound = \frac{18}7, \incumbentValue = 3 $)\\pruned!};
		\node[align=center] (e) at (2*\horidist, -2*\vertdist) {5. ($ \upperbound = \frac{18}7, \incumbentValue = 3 $)\\pruned!};
		\draw (c) edge node[midway,left] {$x_2\ge1 \ $} (d);
		\draw (c) edge node[midway,right] {$ \ x_2\le0$} (e);
	\end{tikzpicture}
\end{adjustbox}

%% file: chapters/protocol.tex
\section{Privacy-Preserving Kidney Exchange using Integer Programming}\label{sec:protocol}

In this section, we present our novel SMPC protocol~$ \kepIpProtocol $ for solving the KEP using a privacy-preserving implementation of the Branch-and-Bound algorithm (cf.~Section~\ref{sec:branch_and_bound}). First, we formally define the ideal functionality~$ \kepIpFunc $ that is implemented by our protocol (Section~\ref{sub:ideal_func}). Then, we describe the setup of the initial tableau for the subset formulation (Section~\ref{sub:tableau_setup}) and provide the specification of our protocol for Branch-and-Bound (Section~\ref{sub:pp_branch_and_bound}). Finally, we provide the complete specification of the protocol~$ \kepIpProtocol $ (Section~\ref{sub:kep_protocol}).

\subsection{Ideal Functionality}\label{sub:ideal_func}

The goal of this paper is to develop a protocol that achieves a superior runtime performance compared to the existing SMPC protocols for solving the KEP~\cite{Breuer_KEprotocol_2020,Breuer_Matching_2022}. 
We achieve this by using the Branch-and-Bound algorithm (cf.~Algorithm~\ref{alg:branch_and_bound}) as the basis for our protocol. 
The performance of the Branch-and-Bound algorithm heavily depends on the pruning of branches in the Branch-and-Bound tree that cannot further improve the best known solution. 
In order to preserve this property of the Branch-and-Bound algorithm in the privacy-preserving setting, we deliberately include the pruning decisions and thus the structure of the Branch-and-Bound tree~$ \branchAndBoundStructure $ in the output of the ideal functionality that is implemented by our novel protocol.

To solve the LP at each node in the tree in a privacy-preserving fashion, we then require a privacy-preserving protocol for the Simplex algorithm.
The efficiency of the Simplex algorithm relies on an early termination if the optimal solution has been found. Thus, similar to the structure of the Branch-and-Bound tree~$ \branchAndBoundStructure $, we also include the number of Simplex iterations~$ \numSimplexIterations $ that are needed to solve each of the subproblems in $ \branchAndBoundStructure $ in the output of the ideal functionality. 
This also comes with the benefit that we can use one of the existing privacy-preserving protocols for Simplex (e.g.,~\cite{Catrina_PPLinearProgramming_2010,Toft_PPLinearProgramming_2009}).

Functionality~\ref{func:kep_ip} formally defines the ideal functionality~$ \kepIpFunc $ implemented by our novel privacy-preserving protocol for solving the KEP. We adopt the compatibility computation between two patient-donor pairs from~\cite{Breuer_KEprotocol_2020} where the bloodtype of donor and patient, the HLA antigens of the donor, and the patient's antibodies against these are considered.

\begin{functionality}[$ \kepIpFunc $ - Solving the KEP based on~IP]\label{func:kep_ip}
	\textit{
		Let all computing peers hold the secret input quotes $ \enc{\inputQuote{i}} = (\enc{\donorbloodvec{i}} $, $ \enc{\antigenvec{i}} $, $ \enc{\patientbloodvec{i}} $, $ \enc{\antibodyvec{i}}) $ of patient-donor pairs $ \partysymbol{i} $ ($ i \in \parties $), where $ \donorbloodvec{i} $ and $ \antigenvec{i} $ are the donor bloodtype and antigen vectors and $ \patientbloodvec{i} $ and $ \antibodyvec{i} $ are the patient bloodtype and antibody vectors as defined in~\cite{Breuer_KEprotocol_2020}. Then, functionality~$ \kepIpFunc $ is given as 
		\begin{equation*}
			((\enc{d_1}, \enc{r_1}), ..., (\enc{d_1}, \enc{r_1}), \branchAndBoundStructure, \numSimplexIterations) \leftarrow \kepIpFunc(\enc{\inputQuote{1}}, ..., \enc{\inputQuote{\numparties}}),
		\end{equation*}  
		where $ \branchAndBoundStructure $ is the structure of the Branch-and-Bound tree, $ \numSimplexIterations $ is the necessary number of Simplex iterations for solving the subproblems in the tree, and $ (d_i, r_i) $ are the indices of the computed donor and recipient for patient-donor pair~$ \partysymbol{i} $ for a set of exchange cycles that maximizes the number of possible transplants between the patient-donor pairs $ \partysymbol{i} $ ($ i \in \parties $). 
	}
\end{functionality}

In Section~\ref{sub:kep_protocol}, we prove that our protocol securely implements functionality~$ \kepIpFunc $ in the semi-honest model. 
Thus, our protocol achieves the same notion of security as the existing SMPC protocols for solving the KEP~\cite{Breuer_KEprotocol_2020,Breuer_Matching_2022}.
However, in contrast to the existing protocols, our protocol not only outputs the exchange partners for each patient-donor pair but also the structure of the Branch-and-Bound tree~$ \branchAndBoundStructure $ and the corresponding number of Simplex iterations~$ \numSimplexIterations $.
This raises the question whether this extended output allows the computing peers to legitimately infer any additional information compared to the existing protocols for solving the KEP.
In this context, we distinguish two categories of information: 

\emph{Sensitive patient-donor information} is information that can be directly linked to the private input and output of individual patient-donor pairs, e.g., their medical data or computed exchange partners.

\emph{Structural information}, in contrast, is information on the underlying instance of the KEP (i.e., the compatibility graph) that cannot be linked to individual patient-donor pairs.

In Appendix~I, we empirically show that the deliberate inclusion of $ \branchAndBoundStructure $ and $ \numSimplexIterations $ in the output of the ideal functionality does not allow for the deriving of any additional sensitive patient-donor information compared to the existing SMPC protocols for solving the KEP~\cite{Breuer_KEprotocol_2020,Breuer_Matching_2022}. However, we also show that it is possible to derive some structural information from~$ \branchAndBoundStructure $ and $ \numSimplexIterations $.

\subsection{Tableau Setup}\label{sub:tableau_setup}

The first step when solving the KEP using the Branch-and-Bound algorithm is the construction of the initial tableau of the LP relaxation for the subset formulation. Recall that $ \subsets $ is the set of all subsets $ \singleSubset_i \subseteq \nodes $ with $ 2 \le \left| \singleSubset_i \right| \le \ell $, and for $ \singleSubset_i \in \subsets $ the set $ \cycles_i = \left\{\singleCycle_{i, 1}, \ldots, \singleCycle_{i, k_i}\right\} $ contains all exchange cycles that consist of the nodes in $ \singleSubset_i $.

\begin{protocol}[t]
	\caption{$ \ipSetupGate $}
	\label{prot:ip_setup}
	\input{./protocols/ip_setup.tex}

\end{protocol}

Protocol~$ \ipSetupGate $ (Protocol~\ref{prot:ip_setup}) computes the initial tableau~$ \enc{\tableau} $ based on the adjacency matrix~$ \enc{\compGraphMatrix} $ which encodes the compatibility graph. It also creates a mapping~$ \enc{\secretSubsetMap} $ such that $ \enc{\secretSubsetMap}(i, j) = 1 $ (with $ i \in \{1, ..., \vert \subsets \vert\} $ and $ j \in \{1, ..., \completeGraphSize\} $) iff all edges of the cycle~$ \singleCycle_{i, j} $ are present in the compatibility graph and $ j $ is the smallest index with this~property. 

First, the initial tableau $ \enc{T} $ is computed based on the subsets $ \subsets $ and the number of patient-donor pairs~$ \numparties $. Then, the computing peers set the first row of the tableau (i.e., the objective coefficients~$ \vert \singleSubset_i \vert $ for $ \singleSubset_i \in \subsets $) based on the adjacency matrix~$ \enc{\compGraphMatrix} $ which encodes the compatibility graph. In particular, they determine for each cycle $ \singleCycle_{i, j} $ ($ i \in \{1, ..., \vert \subsets \vert\} $, $ j \in \{1, ..., \completeGraphSize\} $) whether it exists in the compatibility graph. To this end, they verify that all edges $ (v, w) $ of the cycle $ \singleCycle_{i, j} $ are present in the matrix~$ \enc{\compGraphMatrix} $. For each subset, the first cycle which exists in the compatibility graph is chosen. Finally, the entry $ \enc{\tableau}(1, i) $ is set to the objective coefficient $ \vert \singleSubset_i \vert $ if a cycle of that subset exists in the compatibility graph and to $ \enc{0} $, otherwise. This ensures that only variables for subsets for which there is a cycle in the compatibility graph are used for the optimization.

\begin{protocol}[t]
	\caption{$ \branchAndBoundProtocol $}
	\label{prot:branch_and_bound}
	\input{./protocols/branch_and_bound.tex}

\end{protocol}

\customParagraph{Security.}
The simulator can compute the initial tableau $ T $ as specified in the protocol since this only requires knowledge of information that is publicly available. For the values of $ \simval{\enc{M'}} $, the simulator can just choose random shares from $ \ints_p $ and the call to the protocol $ \minSelGate $ can be simulated by calling the corresponding sub-simulator on  $ \simval{\enc{M'}} $. The output of $ \minSelGate $ can again be simulated by choosing a random share. Finally, the simulator can compute $ \simval{\enc{T}}(1, i) $ as specified in the protocol using the simulated values $ \simval{\enc{\secretSubsetMap}}(i, j) $ and the initial tableau $ T $ which he computed at the beginning. 

\customParagraph{Complexity.}
Protocol~$ \ipSetupGate $ requires $ \vert \subsets \vert \cdot \maxCompleteGraphSize \cdot \cycleBound $ comparisons and $ \vert \subsets \vert $ calls to $ \minSelGate $, where $ \maxCompleteGraphSize = \max_{\singleSubset_i \in \subsets}(\completeGraphSize) $. As all involved loops can be executed in parallel, $ \ipSetupGate $ has round complexity~$ \bigo{\log(\bitsize)} $. The communication complexity is $ \bigo{\maxCompleteGraphSize \cdot \cycleBound \cdot \vert \subsets \vert \cdot \bitsize} $, where $ \bitsize $ is the bit length of the secret values.

\subsection{Privacy-Preserving Branch-and-Bound Protocol}\label{sub:pp_branch_and_bound}

Protocol~\ref{prot:branch_and_bound} contains the specification of our SMPC protocol $ \branchAndBoundProtocol $ for Algorithm~\ref{alg:branch_and_bound}. It receives the initial tableau~$ \enc{\initialTableau} $ for the subset formulation and the upper bound $ \enc{\upperbound} $ as input and returns an optimal solution $ \enc{\incumbent} $ stating for each subset $ \singleSubset_i \in \subsets $ whether it is part of the optimal solution or not.

\customParagraph{Setup Phase.}
First, the trivial solution to choose no subset at all is established, i.e., all entries of $ \enc{\incumbent} $ are set to $ \enc{0} $ which leads to the value $ \enc{\incumbentValue} = \enc{0} $ of the initial solution. Then, the initial problem is added to the list of problems $ L $ that have to be solved. A problem is defined by a tableau~$ \enc{\tableau} $, an upper bound~$ \enc{\upperbound} $ for all solutions in the subtree, the vector~$ \enc{\partialSolutionVector} $ which stores the subsets that have already been added to the solution on the current path in the tree, and the denominator~$ \enc{\denominator} $. 

The main loop of the protocol is then executed until there is no further subproblem in the list $ L $. At the start of each iteration, the first subproblem $ P $ from the list $ L $ is chosen.

\customParagraph{Solution Phase.}
Before solving the subproblem $ P $, the computing peers check whether the upper bound $ \upperbound $ for solutions of $ P $ is larger than the currently best solution $ \incumbentValue $. If this is not the case (i.e., $ t = 0 $), the current subtree is pruned and the next subproblem is chosen. Otherwise, a privacy-preserving protocol for the Simplex algorithm is executed to find a solution for $ P $. To this end, we use the SMPC protocol for Simplex by Toft~\cite{Toft_PPLinearProgramming_2009}. The protocol $ \simplexGate $ returns the optimal solution $ \enc{\simplexSolutionIncomplete} $ with value~$ \enc{\simplexSolutionValue} $ and the common denominator $ \enc{\denominator} $, which is used to avoid fractional solutions (i.e., the actual solution is $ \enc{\frac{\simplexSolutionIncomplete}{\denominator}} $ and has value $ \enc{\frac{\simplexSolutionValue}{\denominator}} $). However, in our protocol the solution returned by Simplex does not consider those subset variables that have already been forced to $ 1 $ in a previous iteration. Therefore, the partial solution $ \enc{\partialSolutionVector} $ (which was stored with the subproblem $P$) is added to the Simplex solution $ \enc{\simplexSolutionIncomplete} $ resulting in the complete solution $ \enc{\simplexSolutionComplete} $ for the current subproblem. 

\begin{protocol}[t]
	\caption{$ \ipUpdateGate $}
	\label{prot:ip_update}
	\input{./protocols/ip_update.tex}

\end{protocol}

\customParagraph{Update Phase.}
In the next step, the protocol $ \ipUpdateGate $ (Protocol~\ref{prot:ip_update}) is used to update the incumbent $ \enc{\incumbent} $ (i.e., the best integer solution found so far) based on the solution $ \enc{\simplexSolutionComplete} $ for the current subproblem. The protocol also derives the index vector $ \enc{\newConstraintVector} $ of the first fractional variable if~$ \enc{\simplexSolutionComplete} $ is fractional.

Instead of dividing the solution~$ \enc\simplexSolutionComplete $ and the optimal value $ \enc{\simplexSolutionValue} $ by the common denominator $ \enc{\denominator} $, we propose a method based on comparisons using the known bounds on $ \simplexSolutionComplete $ and $ \simplexSolutionValue $. Thereby, we avoid fractional values and integer division, which is computationally expensive~\cite{Toft_PPLinearProgramming_2009}. In particular, we exploit the fact that $\frac{\simplexSolutionComplete(i)}{\denominator} \in [0,1]$, which implies that $\frac{\simplexSolutionComplete(i)}{\denominator}$ is fractional iff both $\simplexSolutionComplete(i) > 0$ and $ \simplexSolutionComplete(i) < \denominator $. These two comparisons are then evaluated such that $ \enc{G}(i) = \enc{1} $ iff $\frac{\simplexSolutionComplete(i)}{\denominator}$ is fractional. Afterwards, the index vector~$ \enc{\newConstraintVector} $ of the first fractional entry is determined. 

Finally, the incumbent is updated iff the newly computed solution~$ \enc{\simplexSolutionComplete} $ is better (i.e., if $ \frac{\simplexSolutionValue}{\denominator} > \incumbentValue $) and integral (i.e., $ \newConstraintVector = (0,\ldots,0) $). Note that $ t < t^* $ iff the solution is fractional since $ \numparties \cdot \denominator $ is an upper bound on $ \simplexSolutionValue $. Furthermore, if $ \simplexSolutionComplete(i) \in \left\{0, \denominator\right\} $, then $ \enc{L}(i) = [\simplexSolutionComplete(i) > 0] = \enc{\frac{\simplexSolutionComplete(i)}{\denominator}}$ already encodes the solution $ L = \frac{1}{\denominator} \cdot\simplexSolutionComplete $ with value $ \frac{z}{d} $.

\customParagraph{Branching Phase.}
In this phase, we create two new subproblems based on the new constraint encoded by~$ \enc{\newConstraintVector} $. Note that we create them even if the solution is not fractional. In that case, the new subproblems are pruned when checking the upper bound~$ \upperbound $ against the value $ \incumbentValue $ of the incumbent (Line 9, Protocol~\ref{prot:branch_and_bound}) since $ \upperbound $ will be equal to or less than $ \incumbentValue $.

Protocol~$ \ipBranchGate $ (Protocol~\ref{prot:ip_branch}) creates two tableaux~$ \enc{\tableau_1} $ and $ \enc{\tableau_2} $ for the new subproblems such that the fractional variable indicated by $ \enc{\newConstraintVector} $ is once set to $ \enc{0} $ and once to $ \enc{1} $. Instead of adding a new row to the tableau, we just set all entries of the corresponding column to $ \enc{0} $ and update the bounds accordingly. To this end, we first determine the column $ \enc{D} $ corresponding to the branching variable. Afterwards, the entry $ \enc{D_k} $ is multiplied to the new constraint $ \enc{\newConstraintVector^\top, v} $ for each value $ v \in \{0,1\} $ of the branching variable. The protocol~$ \ipBranchGate $ concludes by subtracting the matrix $ \enc{\Delta} $ from $ \enc{T} $.

\begin{protocol}[t]
	\caption{$ \ipBranchGate $}
	\label{prot:ip_branch}
	\input{./protocols/ip_branch.tex}

\end{protocol}

Note that we never branch on a slack variable as the subset formulation (Equation~\ref{eq:subset_formulation}) ensures that if there is a fractional slack variable, then there is also a subset $ \singleSubset_i \in \subsets $ where the variable $ \subsetVariableIndexed{i} $ is fractional. Since Protocol $ \ipUpdateGate $ takes the fractional variable with the lowest index, it always chooses~$ \subsetVariableIndexed{i} $. Hence, we do not have to handle the special case of branching on a slack variable for our protocol $ \ipBranchGate $.

\customParagraph{Security.}
For the initialization of $ \enc{\partialSolutionVector} $, $ \enc{\incumbent} $, and $ \enc{\incumbentValue} $, the simulator chooses random shares from $ \ints_p $. The calls to $ \simplexGate $, $ \ipUpdateGate $, and $ \ipBranchGate $ can be simulated by calling the respective simulators on random inputs. For the security of the protocol~$ \simplexGate $, we refer to~\cite{Toft_PPLinearProgramming_2009}.
Note that for the simulator of $ \simplexGate $ it has to be assumed that he knows the number of iterations that Simplex requires. 
The security of the protocols~$ \ipUpdateGate $ and $ \ipBranchGate $ is trivial to prove as they only operate on secret values and use secure primitives.
For the computation of the value $ \enc{t} $, which determines whether the current subproblem is pruned or not, the simulator then requires knowledge of the tree structure $ \branchAndBoundStructure $. However, recall that the tree structure is part of the output of the ideal functionality~$ \kepIpFunc $. Hence, the simulator can also infer the tree structure and thus knows in which iterations $ t = 0 $ and in which $ t = 1 $. Based on this knowledge, he can simulate the value $ \simval{\enc{t}} $ (Line~9, Protocol~\ref{prot:branch_and_bound}) such that the reconstruction of $ \enc{t} $ yields the correct result. Thereby, he can also determine which subproblems are added to the list $ L $ and which problems are chosen in which iteration. Finally, the simulator only has to set the output $ \simval{\enc{\incumbent}} $ to the known share of the output $ \enc{\incumbent} $.

\customParagraph{Complexity.}
The complexity of protocol~$ \branchAndBoundProtocol $ is dominated by the call to the protocol~$ \simplexGate $ for each subproblem. According to~\cite{Toft_PPLinearProgramming_2009}, one Simplex iteration requires $ \bigo{\numConstraints \cdot (\numConstraints + \numVariables)} $ multiplications and $ \bigo{\numConstraints + \numVariables} $ comparisons, where $ m $ is the number of constraints and $ n $ the number of variables of the LP. The round complexity of one iteration is $ \bigo{\log\log(\numConstraints)} $. Our protocols $ \ipUpdateGate $ and $ \ipBranchGate $ have round complexity~$ \bigo{\log(\bitsize)} $ and $ \bigo{1} $, respectively, and communication complexity $ \bigo{\numVariables \cdot \bitsize} $ and $ \bigo{\numConstraints \cdot \numVariables} $, respectively. Thus, protocol~$ \branchAndBoundProtocol $ has communication complexity $ \bigo{\numSimplexIterations \cdot (\numConstraints^2 + \numConstraints \numVariables + \numConstraints \bitsize + \numVariables \bitsize)} $ and round complexity $ \bigo{\numSimplexIterations \cdot \log\log(\numConstraints) \log(\bitsize)} $, where $ \numSimplexIterations $ is the overall number of Simplex iterations required for an execution of protocol~$ \branchAndBoundProtocol $. Note that the number of Simplex iterations is $ 2^\numVariables $ in the worst case~\cite{Schrijver_LinearProgramming_1998}. Furthermore, $ m = \numparties $ and $ n = \vert \subsets \vert + \numparties $ for the subset formulation. 

\subsection{Privacy-Preserving Protocol for the KEP}\label{sub:kep_protocol}

\begin{protocol}[t]
	\caption{$ \kepIpProtocol $}
	\label{prot:kep_ip}
	\input{./protocols/kep_ip.tex}

\end{protocol}

Protocol~\ref{prot:kep_ip} contains the specification of our novel privacy-preserving protocol~$ \kepIpProtocol $ that computes the optimal set of exchange cycles for a given set of patient-donor pairs using our protocol~$ \branchAndBoundProtocol $ (Protocol~\ref{prot:branch_and_bound}). Previous to the actual protocol execution each pair $ \partysymbol{i} $ ($ i \in \parties $) secretly shares the relevant medical data of its patient and its donor (as defined in Functionality $ \kepIpFunc $) among the computing peers. After the protocol execution, the computing peers send the shares of the computed exchange partners (indicated by $ \outputDonors(i) $ and $ \outputRecipients(i) $) to each patient-donor pair $ \partysymbol{i} $ ($ i \in \parties $).

\customParagraph{Graph Initialization Phase.} At the start of the protocol, the adjacency matrix $ \enc{\compGraphMatrix} $ is computed such that entry $ \enc{\compGraphMatrix}(i,j) $ encodes the medical compatibility between the donor of patient-donor pair~$ \partysymbol{i} $ and the patient of pair $ \partysymbol{j} $. To this end, we use the protocol~$ \compCheck $ from~\cite{Breuer_KEprotocol_2020} which considers the donor of pair~$ \partysymbol{i} $ as compatible with the patient of pair~$ \partysymbol{j} $ if $ \donorbloodvec{i}(k) = \patientbloodvec{j}(k) = 1 $ for at least one $ k \in \{1, ..., \vert \donorbloodvecNotSpec \vert \} $ and if for all $ k \in \{1, ..., \vert \antibodyvecNotSpec \vert \} $ with $ \antigenvec{i}(k) = 1 $ it holds that $ \antibodyvec{j}(k) \neq 1 $.
Note that even if this compatibility check indicates that two pairs are compatible, the final choice of whether a transplant is carried out lies with medical experts.

Before computing the exchange cycles, we shuffle the adjacency matrix at random to ensure that a node in the compatibility graph cannot be linked to any particular patient-donor pair. 
This also ensures that an entry in the tableaux used in the Branch-and-Bound protocol cannot be linked to a particular patient-donor pair. 
Furthermore, the shuffling ensures unbiasedness in the sense that two pairs with the same input have the same chance to be part of the computed solution (independent of their index). We use the shuffling implementation proposed by Zahur et al.~\cite{Zahur_shuffling_2016} which requires $ \bigo{n^2 \log n} $ communication and $ \bigo{\log n} $ rounds, where $ n $ is the input length of the vector to be shuffled.\footnote{We are aware that there are more efficient implementations for shuffling (e.g.,~\cite{Araki_SecureGraphAnalysis_2021}). We use the version from~\cite{Zahur_shuffling_2016} since it is implemented in MP-SPDZ and shuffling only has a small influence on the runtime of our protocol.}

\customParagraph{Setup Phase.} In the next step, the computed adjacency matrix is used to set up the initial tableau for the subset formulation of the KEP (cf.\ Equation~\ref{eq:subset_formulation}). To this end, we call the protocol~$ \ipSetupGate $ on the shuffled adjacency matrix~$ \enc{\compGraphMatrix} $. Again note that by using the shuffled adjacency matrix, we hide the relation between the rows in the tableau and the indices of the patient-donor pairs. In addition to the secret initial tableau $ \enc{\tableau} $, the protocol~$ \ipSetupGate $ returns the matrix $ \enc{\secretSubsetMap} $ mapping each subset $ \singleSubset_i \in \subsets $ to an exchange cycle (cf. Section~\ref{sub:tableau_setup}). 

\customParagraph{Optimization Phase.} Now we can just call our protocol $ \branchAndBoundProtocol $ on the tableau $ \enc{\tableau} $ using the number of patient-donor pairs $ \numparties $ as upper bound $ \upperbound $. This yields the solution $ \enc{\incumbent} $ with $ \enc{\incumbent}(i) = 1 $ iff the optimal set of exchange cycles contains a cycle with vertex set $ \singleSubset_i $ and $ \enc{\incumbent}(i) = 0 $, otherwise.

\customParagraph{Resolution Phase} Based on the chosen subsets indicated by $ \enc{\incumbent} $, we determine the particular cycles that comprise the optimal solution. To this end, we convert $ \enc{\incumbent} $ into the matrix $ \enc{Y} \in \{0, 1\}^{\numparties \times \numparties} $ such that $ Y(i, j) = 1 $ iff cycle $ \singleCycle_{i, j} $ is part of the optimal set of exchange cycles (i.e., iff $ \incumbent(i) = 1 $ and $\secretSubsetMap(i, j)=1$). Then, we construct the solution matrix $ \solutionMatrix \in \{0, 1\}^{\numparties \times \numparties} $ where $ \enc{\solutionMatrix}(v, w) $ indicates whether the edge~$ (v, w) $ is part of the computed set of exchange cycles. In particular, we compute $ \enc{\solutionMatrix}(v, w) $ as the sum over those entries $ \enc{Y}(i, j) $ that correspond to cycles $ C_{i, j} $ where $ (v, w) \in E(\singleCycle_{i. j}) $. 

\customParagraph{Output Derivation Phase.}
In the last phase, we map the solution matrix to each patient-donor pair's exchange partners. First, the initial shuffling of the pairs is reverted by calling the protocol $ \reverseShuffleGate $ on the solution matrix. As by definition of the KEP each pair can only be involved in at most one exchange cycle, we know that each row and each column of the solution matrix can only contain at most one entry which is equal to $ 1 $. Therefore, we can just add up all entries of row $ i $ multiplied with their column index $ j $ to obtain the index of the pair $ \partysymbol{j} $ whose donor donates to the patient of pair $ \partysymbol{i} $. In a similar fashion we obtain the index of the recipient for the donor of pair $ \partysymbol{i} $ summing up the entries of the $ i $-th column multiplied by their row index~$ j $.

\customParagraph{Security.}
The simulator can compute the adjacency matrix $ \compGraphMatrix $ by calling the simulator of $ \compCheck $ on the corresponding inputs of the patient-donor pairs, simulating the output using a random share. Similarly, he can simulate the calls to the protocols $ \shuffleGate $, $ \ipSetupGate $, $ \branchAndBoundProtocol $, and $ \reverseShuffleGate $. Recall that the simulation of the protocol~$ \branchAndBoundProtocol $ requires knowledge of the structure of the Branch-and bound tree~$ \branchAndBoundStructure $ and the number of Simplex iterations~$ \numSimplexIterations $. However, since this information is part of the output of the ideal functionality~$ \kepIpFunc $, it is available to the simulator. The initialization of the matrix $ A $ can then again be simulated using random shares. Afterwards, the simulator can compute the values for $ \simval{\enc{Y}} $ and $ \simval{\enc{A}} $ as specified in the protocol based on the previously simulated values $ \simval{\enc{\incumbent}} $, $ \simval{\enc{\secretSubsetMap}} $, and $ \simval{\enc{A}} $. Finally, the simulator just has to set the values of the output vectors $ \simval{\enc{\outputDonors}} $ and $ \simval{\enc{\outputRecipients}} $ to the known shares of the protocol output. Thus, protocol~$ \kepIpProtocol $ securely implements functionality~$ \kepIpFunc $.

\customParagraph{Complexity.} 
Communication and round complexity of protocol~$ \kepIpProtocol $ are dominated by the call to the protocol~$ \branchAndBoundProtocol $, i.e., protocol~$ \kepIpProtocol $ has the same complexities as protocol~$ \branchAndBoundProtocol $.

%% file: protocols/ip_setup.tex
\algrenewcommand\algorithmicindent{1em}
\small
\textbf{Input:} Secret adjacency matrix~$ \enc{\compGraphMatrix} $\vspace*{0.2em} \\
\textbf{Output:} Initial tableau~$ \enc{\tableau} $ and mapping $ \enc{\secretSubsetMap} $ from subsets to cycles
\begin{algorithmic}[1]
	\State Compute an initial tableau $ T \in \ints^{(\numparties+1) \times (\vert \subsets \vert + \numparties +1)} $ for Equation~\ref{eq:subset_formulation} such that column $ i $ corresponds to variable $ x_{S_i} $ for $ i \in \{1, ..., \vert \subsets \vert\} $ and to the slack variables for $ i \in \{\vert \subsets \vert + 1, ..., \vert \subsets \vert + \numparties\} $.\vspace*{0.2em}
	\For{$ i = 1, ..., \vert \subsets \vert $}
		\For{$ j = 1, ..., \completeGraphSize $}
			\State $ \enc{M'}(i, j) \leftarrow \sum_{(v, w) \in E(C_{ij})} \enc{\compGraphMatrix}(v, w) \sharedGreaterThan \enc{|S_i| - 1} $
		\EndFor
		\vspace*{0.1em}
		\State $ \enc{\secretSubsetMap}(i) \leftarrow \minSelGate(\enc{M'}(i, 1), ..., \enc{M'}(i, k_i)) $
		\vspace*{0.2em}
		\State $ \enc{T}(1, i) \leftarrow T(1, i) \cdot \sum_{j = 1}^{k_i} \enc{\secretSubsetMap}(i, j) $	
	\EndFor
	
\end{algorithmic}

%% file: protocols/branch_and_bound.tex
\algrenewcommand\algorithmicindent{0.8em}
\small
\textbf{Input:}  Initial tableau $ \enc{\initialTableau} $, initial upper bound $ \enc{\upperbound} $\vspace*{0.2em} \\
\textbf{Output:} Optimal solution $ \enc{\incumbent} $ to the IP represented by $ \enc{\initialTableau} $
\begin{algorithmic}[1]
	\vspace*{0.1em}
	\For{$ i \in \vert \subsets \vert + \numparties $}
		\State $ \enc{\partialSolutionVector}(i) \leftarrow \enc{0} $
		\vspace*{0.1em}
		\State $ \enc{\incumbent}(i) \leftarrow \enc{0} $
	\EndFor
	
	\State $ \enc{\incumbentValue} \leftarrow \enc{0}, \ \enc{\denominator^*} \leftarrow \enc{1} $
	\vspace*{0.1em}
	\State $ L := \{\enc{\initialTableau}, \enc{\upperbound}, \enc{\partialSolutionVector}, \enc{\denominator^*}\} $
	\vspace*{0.1em}
	\While{$ L \neq \emptyset $}
		\vspace*{0.1em}
		\State Choose $ P := (\enc{\tableau}, \enc{\upperbound}, \enc{\partialSolutionVector}, \enc{\denominator}) \in L $ and set $ L := L \setminus \{P\} $
		\vspace*{0.1em}
		\State $ \enc{t} \leftarrow \enc{\upperbound} \cdot \enc{\denominator^*} \sharedGreaterThan \enc{\incumbentValue} \cdot \enc{\denominator} $
		\vspace*{0.1em}
		\State Recover $ \enc{t} $ such that all computing peers learn $ t $
		\vspace*{0.1em}
		\If{$ t = 0 $}
			\vspace*{0.1em}
			\State Go to Step $ 7 $
			\vspace*{0.1em}
		\EndIf
		\vspace*{0.1em}
		\State $ \enc{\simplexSolutionIncomplete}, \enc{\simplexSolutionValue}, \enc{\denominator} \leftarrow \simplexGate(\enc{\tableau}) $
		\vspace*{0.1em}
		\For{$ v \in \nodes $}
			\vspace*{0.1em}
			\State $ \enc{\simplexSolutionComplete}(v) \leftarrow \enc{\simplexSolutionIncomplete}(v) \sharedAdd \enc{\partialSolutionVector}(v) \cdot \enc{\denominator} $
			\vspace*{0.1em}
		\EndFor
		\vspace*{0.1em}
		\State $ \enc{\incumbent}, \enc{\incumbentValue}, \enc{\denominator^*}, \enc{\newConstraintVector} \leftarrow \textsc{\footnotesize{IP-Update}}(\enc{\incumbent}, \enc{\incumbentValue}, \enc{\denominator^*}, \enc{\simplexSolutionComplete}, \enc{\simplexSolutionValue}, \enc{\denominator}) $
		\vspace*{-0.8em}
		\State $ \enc{\tableau_1}, \enc{\tableau_2} \leftarrow \ipBranchGate(\enc{\tableau}, \enc{\newConstraintVector}) $
		\vspace*{0.1em}
		\State $ L := L \cup \{(\enc{\tableau_1}, \enc{z}, \enc{\partialSolutionVector}, \enc{\denominator}), (\enc{\tableau_2}, \enc{z}, \enc{\partialSolutionVector} \sharedAdd \enc{\newConstraintVector}),  \enc{\denominator}\} $
		\vspace*{0.1em}
	\EndWhile
\end{algorithmic}

%% file: protocols/ip_update.tex
\algrenewcommand\algorithmicindent{1em}
\small
\textbf{Input:} Current incumbent $ \enc{\incumbent} $ with value $ \enc{\incumbentValue} $ and denominator $ \enc{\denominator^*} $, solution $ \enc{\simplexSolutionComplete} $ with value $ \enc{\simplexSolutionValue} $ and denominator~$ \enc{\denominator} $ returned by the protocol $ \simplexGate $\vspace*{0.2em}\\		
\textbf{Output:} New incumbent $ \enc{\incumbent} $ with value $ \enc{\incumbentValue} $ and denominator $ \enc{\denominator^*} $, and index vector $ \enc{F} $ of the first fractional entry
\begin{algorithmic}[1]
	\For{$ i = 1, ..., \vert \subsets \vert + \numparties $}
		\State $ \enc{L}(i) \leftarrow \enc{\simplexSolutionComplete}(i) \sharedGreaterThan \enc{0} $
		\vspace*{0.1em}		
		\State $ \enc{H}(i) \leftarrow \enc{\denominator} \sharedGreaterThan \enc{\simplexSolutionComplete}(i) $
		\vspace*{0.1em}		
		\State $ \enc{G}(i) \leftarrow \enc{L}(i) \sharedAdd \enc{H}(i) \sharedSubtract \enc{1} $
		\vspace*{0.1em}		
	\EndFor
	\vspace*{0.1em}		
	\State $ \enc{F} \leftarrow \minSelGate(\enc{G}) $
	\vspace*{0.1em}
	\State $ \enc{t} \leftarrow \enc{\simplexSolutionValue} \cdot \enc{\denominator^*} \sharedSubtract \largestValuePossible \sharedMult \sum_{i=1}^{\vert \subsets \vert + \numparties}\enc{F}(i) $
	\vspace*{0.1em}
	\State $ \enc{t^{\ast}} \leftarrow \enc{\incumbentValue} \sharedMult \enc{\denominator} $
	\vspace*{0.1em}		
	\State $ \enc{\incumbent}, \enc{\incumbentValue}, \enc{\denominator^*} \leftarrow (\enc{t} \sharedGreaterThan \enc{t^{\ast}}) \ ? \ (\enc{L}, \enc{z}, \enc{\denominator})  : (\enc{\incumbent}, \enc{\incumbentValue}, \enc{\denominator^*}) $

\end{algorithmic}

%% file: protocols/ip_branch.tex
\algrenewcommand\algorithmicindent{1em}
\small
\textbf{Input:}  Tableau $ \enc{T} $, vector~$ \enc{F} $ encoding a new constraint\vspace*{0.2em} \\
\textbf{Output:} Two tableaux $ \enc{T^0} $, $ \enc{T^1} $ encoding two new subproblems
\begin{algorithmic}[1]
	\vspace*{0.1em}
	\State $ \enc{D} \leftarrow \enc{T_{-, [1:\vert \subsets \vert + \numparties]}} \sharedMult \enc{F} $
	\vspace*{0.1em}
	\For{$ k = 1, ..., \numparties + 1 $}
		\vspace*{0.1em}
		\State $ \enc{\Delta_k^0} \leftarrow \enc{D}(k) \sharedMult \enc{F^\top, 0} $
		\vspace*{0.1em}
		\State $ \enc{\Delta_k^1} \leftarrow \enc{D}(k) \sharedMult \enc{F^\top, 1} $
		\vspace*{0.1em}
		\State $ \enc{T^0_{k, -}} \leftarrow \enc{T}(k, -) \sharedSubtract \enc{\Delta_k^0} $
		\vspace*{0.1em}
		\State $ \enc{T^1_{k, -}} \leftarrow \enc{T}(k, -) \sharedSubtract \enc{\Delta_k^1} $
		\vspace*{0.1em}
	\EndFor
	
\end{algorithmic}

%% file: protocols/kep_ip.tex
\algrenewcommand\algorithmicindent{1em}
\small
\textbf{Input:}  Each patient-donor pair $ \partysymbol{i} \in \{1, ..., \numparties\} $ shares its medical inputs $ \donorbloodvec{i}, \antigenvec{i}, \patientbloodvec{i}, \antibodyvec{i} $ with the computing peers\vspace*{0.2em} \\
\textbf{Output:} Each computing peer sends the shares of the exchange partners $ \enc{\outputDonors}(i) $ and $ \enc{\outputRecipients}(i) $ to patient-donor pair $ P_i $ with $ i \in \{1, ..., \numparties\} $
\begin{algorithmic}[1]
	\vspace*{0.1em}
	\For{$ i, j \in \{1, ..., \numparties\} $}
		\vspace*{0.1em}
			\State $ \enc{\compGraphMatrix}(i, j) \leftarrow \compCheck(\enc{\donorbloodvec{i}}, \enc{\antigenvec{i}}, \enc{\patientbloodvec{j}}, \enc{\antibodyvec{j}}) $
			\vspace*{0.1em}
	\EndFor
	\vspace*{0.1em}
	\State $ \enc{\compGraphMatrix'} \leftarrow \shuffleGate(\enc{\compGraphMatrix}) $
	\vspace*{0.1em}
	\State $ \enc{\tableau}, \enc{\secretSubsetMap} \leftarrow \ipSetupGate(\enc{\compGraphMatrix'}) $
	\vspace*{0.1em}
	\State $ \enc{\incumbent} \leftarrow \branchAndBoundProtocol(\enc{\tableau}, \enc{\numparties}) $ 
	\vspace*{0.1em}
	\State Let $ \enc{\solutionMatrix} $ be an $ (\numparties \times \numparties) $ adjacency matrix filled with zeros
	\vspace*{0.1em}
	\For{$ i = 1, ..., \vert \mathcal{S} \vert $}
		\vspace*{0.1em}
		\For{$ j = 1, ..., k_i $}
			\vspace*{0.1em}
			\State $ \enc{Y}(i, j) \leftarrow \enc{\incumbent}(i) \sharedMult \enc{\secretSubsetMap}(i, j) $
			\vspace*{0.1em}
			\For{$ (v, w) \in E(\singleCycle_{ij}) $}
			\vspace*{0.1em}
				\State $ \enc{\solutionMatrix}(v, w) \leftarrow \enc{\solutionMatrix}(v, w) \sharedAdd \enc{Y}(i, j) $
				\vspace*{0.1em}
			\EndFor
		\EndFor
	\EndFor
	\vspace*{0.1em}
	\State $ \enc{\solutionMatrix'} \leftarrow \reverseShuffleGate(\enc{\solutionMatrix}) $
	\vspace*{0.1em}
	\For{$ i = 1, ..., \numparties $}
		\vspace*{0.1em}
		\State $ \enc{\outputDonors}(i) \leftarrow \sum_{j=1}^{\numparties}(j \sharedScalMult \enc{\solutionMatrix'}(j, i)) $
		\vspace*{0.1em}
		\State $ \enc{\outputRecipients}(i) \leftarrow \sum_{j=1}^{\numparties}(j \sharedScalMult \enc{\solutionMatrix'}(i, j)) $
		\vspace*{0.1em}	
	\EndFor
\end{algorithmic}

%% file: chapters/evaluation.tex
\section{Evaluation}\label{sec:evaluation}

In this section, we evaluate the performance of our implementation of our protocol~$ \kepIpProtocol $ (Protocol~\ref{prot:kep_ip}). First, we describe the evaluation setup (Section~\ref{sub:eval_setup}). Then, we evaluate runtime and network traffic of our protocol and compare the results to the existing protocol from~\cite{Breuer_KEprotocol_2020} (Section~\ref{sub:eval_runtime}). 

\subsection{Setup}\label{sub:eval_setup}

\begin{table}[t]
	\caption[Runtime plot]{Average runtime and network traffic for our~protocol~\textsc{Kep-IP} and the implementation~\textsc{Kep-Rnd-SS}~\cite{Breuer_Matching_2022} of the protocol from~\cite{Breuer_KEprotocol_2020}.}
	\label{tab:runtimes}
	\input{tables/runtimes_compared_latencies.tex}
\end{table}

We have implemented our protocol~$ \kepIpProtocol $ using the SMPC benchmarking framework MP-SPDZ~\cite{KellerMPSPDZ2020} version 0.3.5. We use the replicated secret sharing scheme from~\cite{Araki_ReplicatedSecretSharing_2016}, which to the best of our knowledge to date is the most efficient scheme for three computing peers with semi-honest security and an honest majority.\footnote{In the original version~\cite{Breuer_KepIp_2022} of this paper we used Shamir secret sharing~\cite{ShamirSecretSharing1979} and MP-SPDZ version 0.2.5. Besides these changes, for this extended version of the paper we were also able to parallelize a larger part of the protocol~$ \textsc{Kep-IP} $ in MP-SPDZ. Together with the better performance of replicated secret sharing compared to Shamir secret sharing, this leads to a significant runtime improvement compared to the runtimes presented in the original version of this paper.} 
We adopt the setup from~\cite{Breuer_Matching_2022}, where different containers on a single server with an AMD EPYC 7702P 64-core processor are used for computing peers and input peers. Each container runs Ubuntu 20.04 LTS, has a single core, and 4GB RAM. In our evaluations, we then run each of the three computing peers on a separate container. A fourth container is used to manage the sending of input and the receiving of output for all patient-donor pairs. Similar to~\cite{Breuer_Matching_2022}, we evaluate our protocol for three different values for the latency $ L $ (i.e., 1ms, 5ms, and 10ms) and a fixed bandwidth of 1Gbps between the computing peers.

In order to provide realistic values for the medical data of the patient-donor pairs, we use a real-world data set from the United Network for Organ Sharing (UNOS) which is a large kidney exchange platform in the US.~\footnote{The data reported here have been supplied by the United Network for Organ Sharing as the contractor for the Organ Procurement and Transplantation Network. The interpretation and reporting of these data are the responsibility of the author(s) and in no way should be seen as an official policy of or interpretation by the OPTN or the U.S.\ Government.} The data set contains all patient-donor pairs that registered with the platform between October 27th 2010 and December 29th 2020 yielding a total of 2913 unique patient-donor pairs from which we sample uniformly at random for each protocol execution.

\subsection{Runtime Evaluation}\label{sub:eval_runtime}

Table~\ref{tab:runtimes} shows runtime and network traffic of our protocol~$ \kepIpProtocol $ and the protocol~$ \breuerSecSharingProt $ from~\cite{Breuer_Matching_2022}, which is a more efficient implementation of the protocol from~\cite{Breuer_KEprotocol_2020}. For both protocols, we consider a maximum cycle size $ \cycleBound = 3 $ as this is the most common maximum cycle size used in existing kidney exchange platforms~\cite{Biro_EuropeanKE_2019,AshlagiKidneyExchangeOperations2021}. The results for our protocol~$ \kepIpProtocol $ are averaged over 50 repetitions for each value of $ \numparties $ (number of patient-donor pairs). Due to the large amount of repetitions required for each value of $ \numparties $, we evaluate the protocol~$ \kepIpProtocol $ for an average runtime up to two hours.

\begin{figure}[t]
	\includegraphics[width=\columnwidth]{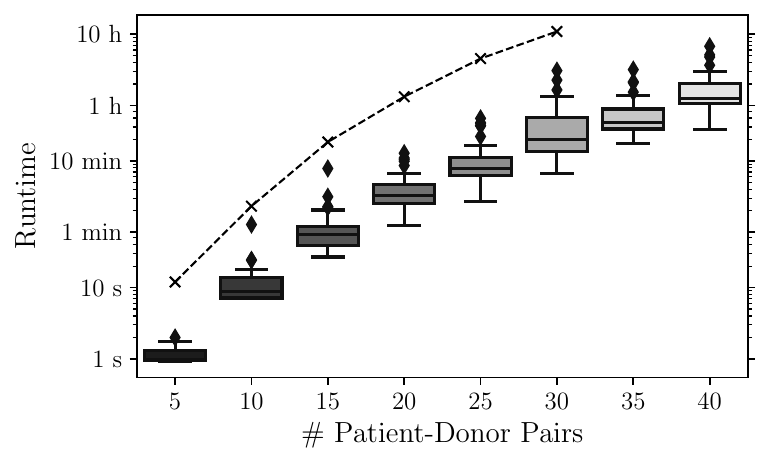}
	\vspace*{-1.5em}
	\caption{Boxplots for the runtime of protocol~$ \textsc{Kep-IP} $ for L = 1ms. The dashed line plot shows the average runtimes from the original version of this paper, where the protocol was evaluated using Shamir Secret Sharing and MP-SPDZ version 0.2.5 and where less parallelization was used.}
	\label{fig:runtimes}
\end{figure}

When comparing the performance of our protocol $ \kepIpProtocol $ to the protocol~$ \breuerSecSharingProt $, we observe that the runtime of both protocols is about the same for up to $ 8 $ patient-donor pairs. However, for larger numbers of pairs our protocol significantly outperforms the protocol~$ \breuerSecSharingProt $. For example, for 12 pairs our protocol is already more than 500 times faster. This is due to the brute force nature of protocol~$ \breuerSecSharingProt $ which scales in the number of potential exchange constellations between the involved patient-donor pairs (cf.~Section~\ref{sub:rw_pp_ke}). While this number is still very low for small values of $ \numparties $, it quickly increases for larger numbers of pairs. 

Furthermore, we observe that the runtime of our protocol~$ \kepIpProtocol $ increases with increasing values for the latency~$ L $. The factor between the runtimes for the different latencies is nearly constant which is to be expected as the latency is just a constant delay for each message. In particular, the runtimes for $ L = 5ms $ and $ L = 10ms $ are about $ 4.31 $ times and $ 8.28 $ times larger than for $ L = 1ms $, respectively.

Another important observation is that the runtime of our protocol varies for different inputs. Figure~\ref{fig:runtimes} shows boxplots for latency $ L = 1ms $. We observe that there are outliers where our protocol performs significantly worse than the average. This is due to the nature of the Branch-and-Bound algorithm and the Simplex algorithm which require different numbers of subproblems and iterations depending on the input. However, even for the worst case which we measured in our evaluations, our protocol outperforms the protocol~$ \breuerSecSharingProt $ from~\cite{Breuer_Matching_2022}. 

Figure~\ref{fig:runtimes} also shows that the runtime of our protocol increases sub-exponentially on average. Thus, we can assume that the runtimes for more than 40 patient-donor pairs will still be feasible for the use case of kidney exchange where a match is usually computed once every couple of days or even only once every 3 months~\cite{Biro_EuropeanKE_2019}.

%% file: tables/runtimes_compared_latencies.tex
\begin{adjustbox}{width=\columnwidth}
	\centering
	\begin{tabular}{c | r r r r | r r}
		\multirow{1}{*}{patient-} 	& \multicolumn{4}{c |}{\textsc{Kep-IP}}	& \multicolumn{2}{c}{\textsc{Kep-Rnd-SS}~\cite{Breuer_Matching_2022}}	\\ \cline{2-7}
		\multirow{1}{*}{donor}	& \multicolumn{3}{c |}{runtime [s]}	& \multicolumn{1}{c |}{\multirow{1}{*}{traffic}}	& \multicolumn{1}{c |}{runtime [s]}	& \multicolumn{1}{c}{\multirow{1}{*}{traffic}} \Bstrut\Tstrut\\ \cline{2-4} \cline{6-6}
		\multirow{1}{*}{pairs}	& L = 1ms & L = 5ms & \multicolumn{1}{r |}{L = 10ms}  & \multicolumn{1}{c |}{\multirow{1}{*}{[MB]}}								& 	\multicolumn{1}{r |}{L = 1ms} 						& 		\multicolumn{1}{c}{\multirow{1}{*}{[MB]}}							\Bstrut\Tstrut\\ \hline
		3 	& $<$1 	& 2		& 3		& 1		& $<$1	& 1\Tstrut\\
		4 	& $<$1	& 3		& 6		& 2		& $<$1	& 2		\\
		5 	& 1		& 5		& 9		& 4		& 1		& 4		\\
		6 	& 2		& 8		& 16	& 8		& 2		& 9		\\
		7 	& 3		& 14	& 28	& 14 	& 3		& 28	\\
		8 	& 5		& 22	& 43	& 21	& 11	& 101	\\
		9 	& 9		& 35	& 68	& 35	& 48	& 398	\\
		10 	& 12	& 51	& 94	& 49	& 326	& 3069	\\
		11 	& 19	& 81	& 157	& 76	& 1649	& 12711	\\
		12 	& 23	& 97	& 188	& 95	& 12087	& 100034\\
		13 	& 34	& 141	& 270	& 137	& -		& -		\\
		14 	& 46	& 205	& 385	& 188	& -		& -		\\
		15 	& 70	& 303	& 556	& 270	& -		& -		\\
		20 	& 247	& 1169	& 2232	& 952	& -		& -		\\
		25 	& 653	& 2902	& 5682	& 2467 	& -		& -		\\
		30 	& 1983	& 8967	& -		& 7442 	& -		& -		\\
		35 	& 2802	& -		& -		& 10906	& -		& -		\\
		40 	& 6303	& -		& -		& 24119	& -		& -		\\
	\end{tabular}
\end{adjustbox}

%% file: chapters/relatedWork.tex
\section{Related Work}

In this section, we review relevant related work including SMPC protocols for kidney exchange (Section~\ref{sub:rw_pp_ke}) and LP (Section~\ref{sub:rw_pp_lp}) as well as privacy-preserving approaches for distributed constraint optimization~(Section~\ref{rw:pp_dcop}).

\subsection{Privacy-Preserving Kidney Exchange}\label{sub:rw_pp_ke}

The first SMPC protocol for solving the KEP was developed by Breuer et al.~\cite{Breuer_KEprotocol_2020} based on homomorphic encryption, which was improved upon in~\cite{Breuer_Matching_2022} with a more efficient implementation based on secret sharing. This protocol uses a pre-computed set of all possible constellations of exchange cycles that can exist between a set of patient-donor pairs. For each of these constellations it then determines whether the contained cycles also exist in the compatibility graph induced by the input of the patient-donor pairs. Finally, the protocol chooses one of those sets that maximize the number of patients that can receive a transplant uniformly at random.
The runtime of the protocol increases quickly for an increasing number of patient-donor pairs due to the brute force nature of the approach. In contrast, our protocol $ \kepIpProtocol $ (Section~\ref{sec:protocol}) uses IP as the underlying technique which leads to a significantly better performance.

In a second line of work, Breuer et al.~\cite{Breuer_Matching_2022} present a dedicated protocol for crossover exchange (i.e., exchange cycles of size $ 2 $) that computes a maximum matching on general graphs. In contrast, our protocol~$ \kepIpProtocol $ allows for the specification of an arbitrary upper bound $ \cycleBound $ on the maximum cycle size.

In comparison to~\cite{Breuer_KEprotocol_2020,Breuer_Matching_2022}, the output of the ideal functionality that is implemented by our protocol~$ \kepIpProtocol $ includes not only a set of exchange cycles that maximizes the number of possible transplants but also the structure of the Branch-and-Bound tree and the number of Simplex iterations necessary to solve each subproblem in the tree.

Recently, Birka et al.~\cite{Birka_PPKidneyExchange_2022} present another SMPC protocol in the context of kidney exchange.  In contrast to the protocols from~\cite{Breuer_KEprotocol_2020,Breuer_Matching_2022} as well as the newly-developed protocol in this paper, the protocol introduced in~\cite{Birka_PPKidneyExchange_2022} only computes an approximate solution to the KEP. In particular, their protocol greedily computes a set of possible exchanges for a single fixed cycle size~$ \cycleBound $.
This set is only an approximation to a solution of the KEP since it neither maximizes the number of exchange cycles for the fixed~$ \cycleBound $ nor does it consider all possible cycle sizes up to~$ \cycleBound $.
Similar to our novel protocol~$ \kepIpProtocol $, the protocol from~\cite{Birka_PPKidneyExchange_2022} includes certain additional information as part of the protocol output (i.e., the number of exchange cycles that exist in the private compatibility graph). While, similar to our protocol, it is unlikely that this allows for the derivation of any sensitive patient-donor information (i.e., information that can be linked to a particular patient-donor pair), the number of cycles in the graph may allow for the derivation of some structural information (e.g., the size of the optimal solution or the density of the compatibility graph). However, this is not analyzed in~\cite{Birka_PPKidneyExchange_2022}.
While~\cite{Birka_PPKidneyExchange_2022} exhibits a superior runtime performance compared to the protocols from \cite{Breuer_Matching_2022,Breuer_KEprotocol_2020} and our novel protocol~$ \kepIpProtocol $ (cf.~Protocol~\ref{prot:kep_ip}), it fails to analyze the quality of the approximation. This is highly problematic for the use case of kidney exchange where any possible exchange that is not found may have severe ramifications for the respective patient-donor pairs. 

\subsection{Privacy-Preserving Linear Programming}\label{sub:rw_pp_lp}

In privacy-preserving LP, there are transformation-based approaches where constraint matrix and objective function are hidden using a monomial matrix (e.g.,~\cite{Dreier_PPLinearProgramming_2011,Vaidya_PPLinearProgramming_2009}) and SMPC-based approaches where the whole LP solver is implemented as an SMPC protocol~\cite{Catrina_PPLinearProgramming_2010,Li_PPLinearProgramming,Toft_PPLinearProgramming_2009}. As the transformation-based approaches cannot guarantee security in the cryptographic sense that we require for our use case, we do not further discuss them and focus on SMPC-based approaches. 

To the best of our knowledge, the only LP solver for which there are privacy-preserving implementations is the Simplex algorithm.
Li and Atallah~\cite{Li_PPLinearProgramming} propose a two-party protocol for a version of Simplex without divisions. However, in their protocol a single Simplex iteration can double the bitsize of the values in the tableau making their protocol only feasible for problems where the required number of Simplex iterations is small. Toft~\cite{Toft_PPLinearProgramming_2009} devises an SMPC protocol for Simplex based on secret sharing which uses integer pivoting to make sure that the protocol can be run using secure integer arithmetic. Finally, Catrina and de Hoogh~\cite{Catrina_PPLinearProgramming_2010} propose a more efficient protocol for Simplex relying on secure fixed-point arithmetic. 

For our protocol for Branch-and-Bound (Section~\ref{sub:pp_branch_and_bound}), we require a privacy-preserving protocol for Simplex that uses integer arithmetic. Therefore, we use the approach by Toft~\cite{Toft_PPLinearProgramming_2009}. However, any SMPC protocol for Simplex that uses integer arithmetic can be used in our protocol.

\subsection{Privacy-Preserving Distributed Constraint Optimization}\label{rw:pp_dcop}

A related field to IP is Distributed Constraint Optimization (DCOP) where the goal is to find an assignment from a set of variables (each held by a different agent) to a set of values such that a set of constraints is satisfied. There are some privacy-preserving approaches to DCOP which use an adaptation of the Branch-and-Bound algorithm (e.g.,~\cite{Grishpoun_PPDCOP_2016,Tassa_PPDCOP_2021}). However, these approaches cannot be applied to our problem, as DCOP is different from the KEP in that each agent holds a variable and knows its value whereas in the KEP the values of the variables in the IP have to remain secret to all patient-donor pairs as they indicate which exchange cycles are chosen.

%% file: chapters/conclusion.tex
\section{Conclusion and Future Work}

We have developed a novel privacy-preserving protocol for solving the KEP using IP which is the most efficient method for solving the KEP in the non-privacy-preserving setting. We have implemented and evaluated our protocol and shown that it significantly outperforms the existing privacy-preserving protocol for solving the KEP.

While we have evaluated our protocol for a fixed set of patient-donor pairs, existing kidney exchange platforms are dynamic in that the pairs arrive at and depart from the platform over time. In future work, we plan to evaluate the efficiency of our protocol when used in such a dynamic setup. Another interesting direction for future research is the inclusion of \emph{altruistic donors} who are not tied to any particular patient.

%% file: chapters/appendix.tex
\section*{Appendix I. Impact of the Extended Output}\label{app:analysis}
As described in Section~\ref{sub:ideal_func}, we deliberately include the number of Simplex iterations~$ \numSimplexIterations $ and the structure of the Branch-and-Bound tree~$ \branchAndBoundStructure $ in the output of the ideal functionality that is implemented by our protocol~$ \kepIpProtocol $ in order to increase the runtime performance of our protocol. 
We have shown that our protocol provides for security in the same setting as the existing SMPC protocols for solving the KEP~\cite{Breuer_KEprotocol_2020,Breuer_Matching_2022}, which do not include $ \numSimplexIterations $ and $ \branchAndBoundStructure $ in their output (cf.~Section~\ref{sub:kep_protocol}).
In this section, we analyze whether and if so how much information can legitimately be derived from the extended output of the protocol~$ \kepIpProtocol $ in comparison to the output of the protocols in~\cite{Breuer_KEprotocol_2020,Breuer_Matching_2022}, which only provide the suggested exchange partners for the patient-donor pairs.
Recall that we distinguish between \emph{sensitive patient-donor information} which can be directly linked to particular patient-donor pairs and \emph{structural information} which concerns the underlying instance of the KEP but cannot be linked to particular patient-donor pairs (cf.~Section~\ref{sub:ideal_func}). 

\subsection*{A. Algorithmic Observations}
Based on the definition of the Simplex algorithm and the Branch-and-Bound algorithm~(cf.~Algorithm~\ref{alg:branch_and_bound}), it is possible to derive some structural information.
In particular, the Simplex algorithm by definition considers exactly one variable in each iteration~\cite{Schrijver_LinearProgramming_1998}. 
Recall that in the subset formulation of the KEP (cf.~Equation~\ref{eq:subset_formulation}), a variable corresponds to a subset of patient-donor pairs that can form an exchange cycle.
Thus, in our protocol at most one exchange cycle can be added per Simplex iteration. 
Since this is a feature of the Simplex algorithm, it holds irrespective of the underlying problem instance and the used data set.

This relation between the number of Simplex iterations~$ \numSimplexIterations $ and the number of cycles in the optimal solution directly implies that including $ \numSimplexIterations $ in the output of the ideal functionality allows for the deriving of structural information on the underlying instance of the KEP. 
Specifically, if the number of Simplex iterations is equal to $ 0 $, no exchange cycle has been selected at all.
Furthermore, if the number of Simplex iterations is larger than $ 0 $, there is a solution containing at least one exchange cycle.
In general, the number of Simplex iterations required for solving a subproblem in the Branch-and-Bound tree~$ \branchAndBoundStructure $ forms an upper bound on the number of exchange cycles in the optimal solution computed by our protocol~$ \kepIpProtocol $.
However, this does not imply that the exact number of exchange cycles in the optimal solution can be predicted easily based on the number of Simplex iterations, as shown in our empirical evaluation (cf.~Appendix I.B).

The Branch-and-Bound algorithm (cf.~Algorithm~\ref{alg:branch_and_bound}) by definition only requires more than one subproblem if the Simplex algorithm returns a fractional solution for the initial LP of the subset formulation.
This can not occur if there is only a single exchange cycle (either of size $ 2 $ or $ 3 $) in the compatibility graph or if there are only two intersecting cycles of size $ 2 $.
Thus, whenever a second subproblem is required, the optimal value of the solution is at least three.

Overall, by definition of the underlying algorithms it is possible to derive an upper bound on the number of exchange cycles in the optimal solution from~$ \numSimplexIterations $ and a lower bound on the value of the optimal solution if $ \branchAndBoundStructure $ contains more than one subproblem. 
Note, however, that these bounds are not very tight for the instances of the KEP found in practice, as shown in our empirical evaluation (cf.~Appendix~I.B).

\subsection*{B. Empirical Observations}

It is not possible to formally prove that it is generally impossible to derive any sensitive patient-donor information or structural information beyond the bounds determined in Appendix~I.A from the number of Simplex iterations~$ \numSimplexIterations $ and the structure of the Branch-and-Bound tree~$ \branchAndBoundStructure $. 
Therefore, in this section we empirically analyze whether such derivation is possible for the instances of the KEP that occur in a real-world setting.

\customParagraph{Methodology.}
In order to obtain results that reflect the application of our protocol for kidney exchange in practice, we require a data set that resembles the inputs of patient-donor pairs in a real-world setting. 
We use a real-world data set from UNOS (cf.~Section~\ref{sub:eval_setup}) which contains the medical data of 2,913 unique patient-donor pairs that registered with UNOS between October 27th 2010 and December 29th 2020. 
Specifically, in our analysis we use the data of real patient-donor pairs that participated in a kidney exchange platform in the past instead of using a graph generator for kidney exchange (e.g., as in~\cite{Delorme_KEGraphGenerator_2022,Saidman_KEGraphGenerator_2006}) that is only based on statistical distributions observed in the real-world data.

For each instance of the KEP that we evaluate in the following, we sample the patient-donor pairs uniformly at random from all pairs in the UNOS data set. This may yield constellations where a KEP instance includes patient-donor pairs that were not registered with UNOS at the same time. However, this is not problematic since the inputs still resemble the medical data of pairs that participated in a real-world kidney exchange platform.
Note that we cannot simply repeat the match runs from UNOS as these include around $ 200 $ patient-donor pairs on average and thus our protocol would not yield feasible run times for these.

For our evaluations, we then determine the number of Simplex iterations~$ \numSimplexIterations $ and the structure of the Branch-and-Bound tree~$ \branchAndBoundStructure $ for a large number of executions of our protocol~$ \kepIpProtocol $ on the input of patient-donor pairs from the UNOS data set.

In order to be able to run our protocol for a large number of repetitions in a feasible amount of time, we re-implemented our protocol as a non-privacy-preserving algorithm, which does not use any SMPC techniques. 
The number of Simplex iterations~$ \numSimplexIterations $ and the structure of the Branch-and-Bound tree~$ \branchAndBoundStructure $ as well as the results obtained by this non-privacy-preserving implementation are exactly the same as for our protocol~$ \kepIpProtocol $ if executed on the same set of patient-donor pairs.

We evaluate this algorithm for up to $ 40 $ patient-donor pairs, which is the same upper bound as in our run time evaluation (cf.~Section~\ref{sub:eval_runtime}). 
Furthermore, we consider a step size of~$ 5 $ which is sufficient to observe the behavior for an increasing number of pairs. 
For each number of pairs, we then execute the non-privacy-preserving algorithm on 10,000 instances of the KEP.
For each such instance, we choose the patient-donor pairs uniformly at random from the UNOS data set.

\begin{figure}
	\includegraphics[width=\columnwidth]{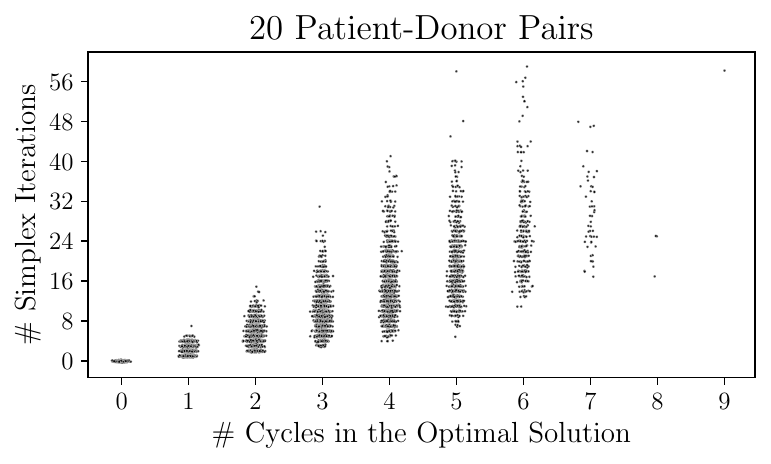}
	\vspace*{-2em}
	\caption{Relationship between the number of Simplex iterations for each node in the Branch-and-Bound tree~$ \branchAndBoundStructure $ that is included in the output of the ideal functionality implemented by our protocol~$ \textsc{Kep-IP} $ and the number of cycles that are included in the corresponding computed optimal solution for 10,000 instances of the KEP for $ 20 $ patient-donor pairs. Each dot corresponds to the number of Simplex iterations for one execution of the Simplex algorithm.}
	\label{plt:correlation_simplex_plt}
\end{figure}

\customParagraph{Number of Simplex Iterations.}
Figure~\ref{plt:correlation_simplex_plt} shows the relationship between the number of cycles in the optimal solution and the number of Simplex iterations for 10,000 instances of the KEP for $ 20 $ patient-donor pairs. 
Each dot in Figure~\ref{plt:correlation_simplex_plt} corresponds to the number of Simplex iterations for one execution of the Simplex algorithm.
Note that the results are similar for any number of patient-donor pairs. 
For completeness, we provide the additional results in Appendix~III.

We observe that the number of cycles in the optimal solution increases with the number of Simplex iterations.
Thus, the number of Simplex iterations is clearly correlated with the number of exchange cycles in the optimal solution and thus also with the size of the optimal solution, i.e., the number of patient-donor pairs that are matched. 
This goes in line with the algorithmic observation that in each iteration at most one exchange cycle can be added to the solution.
However, Figure~\ref{plt:correlation_simplex_plt} also shows that aside from the special cases of $ 0 $ and~$ 1 $ Simplex iterations, it is only possible to make predictions on the number of cycles in the optimal solution with some probability. 
For example, if there are $ 10 $ Simplex iterations, our empirical results suggest probabilities of $ 10\% $ for two cycles, $ 56\% $ for three cycles, $ 29\% $ for four cycles, and $ 4\% $ for five cycles in the optimal solution. 

The correlation between the number of cycles in the optimal solution and the number of Simplex iterations of course also holds for other properties of the problem instance that are correlated with the size of the optimal solution, e.g., the overall number of cycles that exist in the compatibility graph. 
While one might suspect that information such as the existence of a large number of cycles in the optimal solution might indicate that certain medical characteristics such as a particular blood type occur more frequently among the patient-donor pairs than for instances of the KEP with less cycles in the optimal solution, we were not able to find any correlation between the number of Simplex iterations and the distribution of blood types, HLA antigens, or antibodies.

Overall, our measurements empirically show that our decision to include the number of Simplex iterations~$ \numSimplexIterations $ in the output of the ideal functionality allows for the derivation of some structural information whereas it does not allow for the derivation of any sensitive patient-donor information. 
In particular, our empirical analysis suggests that it is highly unlikely to predict the existence of any particular edge in the compatibility graph since the number of Simplex iterations neither indicates which particular exchange cycle is chosen nor which particular patient-donor pairs are part of the computed exchange cycles. 
In addition, this information is obfuscated by shuffling the nodes of the compatibility graph prior to the execution of the Branch-and-Bound protocol. 
Therefore, we can conclude that it is unlikely to infer any sensitive patient-donor information from the number of Simplex iterations.

For the use case of kidney exchange, one may argue that the structural information that can be derived from the number of Simplex iterations is acceptable since certain information such as the number of transplants that are found may be published anyway for statistical purposes. However, in case this is deemed unacceptable, it is possible to obfuscate the information on the number of Simplex iterations~$ \numSimplexIterations $ in the output of the ideal functionality. In Appendix~II, we introduce a possible strategy for obfuscating the number of Simplex iterations and evaluate its effectiveness for the use case of kidney exchange.

\customParagraph{Structure of the Branch-and-Bound Tree.}
In order to analyze whether any further information can be derived from the structure of the Branch-and-Bound tree~$ \branchAndBoundStructure $, we evaluate whether and if so which kind of correlation exists between the number of subproblems (i.e., the size of the tree) and any other information on the problem instance at large.

\begin{figure}
	\includegraphics[width=\columnwidth]{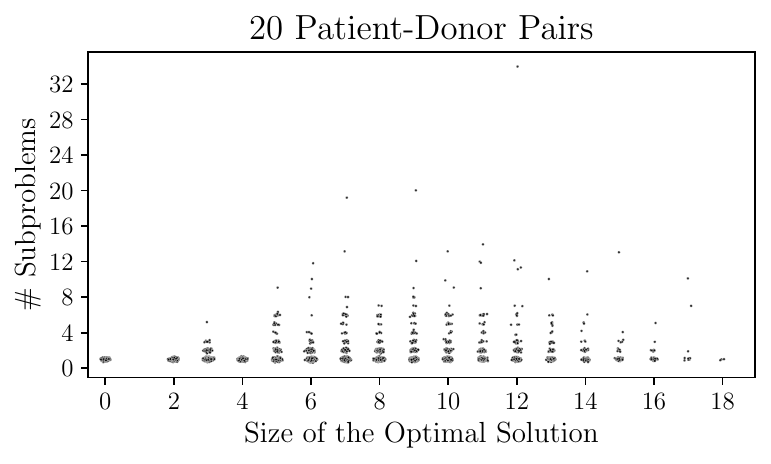}
	\vspace*{-2em}
	\caption{Relationship between the number of subproblems in the Branch-and-Bound tree~$ \branchAndBoundStructure $ for the protocol~$ \textsc{Kep-IP} $ and the size of the optimal solution for 10,000 instances of the KEP for $ 20 $ patient-donor pairs. Each dot corresponds to the number of subproblems for one execution of the Branch-and-Bound algorithm.}
	\label{plt:correlation_bab_plt}
\end{figure}

Figure~\ref{plt:correlation_bab_plt} shows the relationship between the number of subproblems in the Branch-and-Bound tree~$ \branchAndBoundStructure $ and the size of the optimal solution, i.e., the number of transplants in the optimal solution for $ 20 $ patient-donor pairs. 
Note that the results are similar for all numbers of patient-donor pairs. 
For completeness, we provide the additional results in Appendix~III.

Figure~\ref{plt:correlation_bab_plt} confirms that the size of the optimal solution is at least three if more than one subproblem is required.
Aside from this, there is no obvious correlation between the number of subproblems and the size of the optimal solution. 
In particular, larger numbers of subproblems do not correlate with larger numbers of potential transplants. 

In addition to the size of the optimal solution, we also evaluated whether there are correlations between $ \branchAndBoundStructure $ and other structural information such as the number of cycles in the compatibility graph, the number of overlapping cycles, the number of chosen cycles in the optimal solution, the number of patients or donors with a certain blood type, or the average number of antibodies of the participating patients.
However, in contrast to the number of Simplex iterations, we were not able to determine any further correlation between the number of subproblems and any structural information. 
We provide results for the relationships between the number of subproblems and further structural information in Appendix~III.

Overall, our empirical analysis suggests that it is highly unlikely to infer any structural information on the underlying problem instance from the structure of the Branch-and-Bound tree~$ \branchAndBoundStructure $ other than that the solution is at least of size three if more than one subproblem is required. 
Similar to the number of Simplex iterations, our empirical analysis also suggests that it is highly unlikely to infer any sensitive patient-donor information from~$ \branchAndBoundStructure $. 
Still, in Appendix~II we introduce a strategy for obfuscating the information that can be derived from $ \branchAndBoundStructure $ and we evaluate its effectiveness for the use case of kidney exchange.

\section*{Appendix II. Obfuscating the Extended Output}\label{app:limit_leakage}

In Appendix~I, we show that our decision to include the number of Simplex iterations~$ \numSimplexIterations $ and the structure of the Branch-and-Bound tree~$ \branchAndBoundStructure $ in the output of the ideal functionality that is implemented by our protocol~$ \kepIpProtocol $ may allow for the deriving of some structural information. 
In this section, we evaluate in how far we can obfuscate the structural information that can be derived from~$ \numSimplexIterations $ and $ \branchAndBoundStructure $ by providing less detail on $ \numSimplexIterations $ and $ \branchAndBoundStructure $ in the output of the ideal functionality.
First, we determine suitable strategies for obfuscating the information that can be derived from $ \numSimplexIterations $ and $ \branchAndBoundStructure $ based on our empirical results from Appendix~I. 
Then, we evaluate the run time overhead induced by these strategies when applied to our protocol~$ \kepIpProtocol $.

\begin{figure}[t]
	\includegraphics[width=\columnwidth]{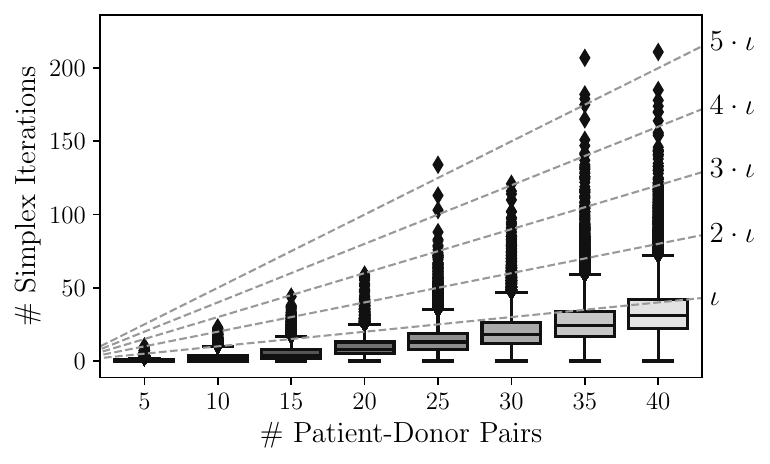}
	\vspace*{-2em}
	\caption{Boxplots for the number of Simplex iterations required for solving the subset formulation of the KEP with our protocol~$ \textsc{Kep-IP} $ for 10000 compatibility graphs where the patient-donor pairs are drawn uniformly at random from the UNOS data set (cf.~Section~\ref{sub:eval_setup}).}
	\label{plt:boxplots_simplex_iterations}
\end{figure}

\customParagraph{Obfuscating the Number of Simplex Iterations.}
It is possible to obfuscate the exact number of Simplex iterations following Toft~\cite{Toft_PPLinearProgramming_2009}, who suggests to obfuscate the number of iterations by only checking for termination every $ \simplexFrequency \in \nats $ iterations. 
Thereby, the output of the ideal functionality no longer contains the exact number of Simplex iterations that are required for solving each of the subproblems in $ \branchAndBoundStructure $. 
Note that the Simplex protocol from~\cite{Toft_PPLinearProgramming_2009} (which we use as a subprotocol in our protocol~$ \kepIpProtocol $) can be implemented such that the solution is not further changed if additional iterations are performed although the optimal solution has already been found. 
Thus, it is straight-forward to apply this technique for obfuscating the exact number of Simplex iterations to the protocol~$ \kepIpProtocol $. 

What remains to be shown is the effectiveness of this approach. 
The effectiveness of course depends on the value of the parameter $ \simplexFrequency $, i.e., on the frequency at which we check the termination condition. 
While a large value for $ \simplexFrequency $ better obfuscates the exact number of iterations, it also implies a larger runtime for the protocol~$ \kepIpProtocol $. 
A commonly stated result for the Simplex algorithm is that it terminates within $ 2m $ or $ 3m $ iterations for most practical problems, where $ m $ is the number of constraints (i.e., $ m = \numparties $ for the subset formulation)~\cite{Nocedal_NumericalOptimization_1999,Quandt_SimplexAverageTermination_1964}.  
This would suggest $ \simplexFrequency = 2\numparties $ and $ \simplexFrequency = 3\numparties $ as suitable values for obfuscating the number of Simplex iterations when using our protocol~$ \kepIpProtocol $ to solve the subset formulation of the KEP.

In order to verify whether this general result also holds for the use case of kidney exchange in practice, we use the results of our empirical evaluation from Appendix~I, where we determined the number of Simplex iterations required by our protocol~$ \kepIpProtocol $ for different numbers of patient-donor pairs.

Figure~\ref{plt:boxplots_simplex_iterations} shows boxplots for the number of Simplex iterations for different numbers of patient-donor pairs. Recall that for each number of pairs we evaluated 10,000 instances of the KEP drawing the inputs of the patient-donor pairs uniformly at random from the real-world data set from UNOS.
The dashed lines indicate different possibilities for choosing the value for $ \simplexFrequency $ depending on the number of patient-donor pairs.
We observe that the majority of the subproblems finish within $ \numparties $ iterations, where $ \numparties $ corresponds to the number of patient-donor pairs in the compatibility graph.
While the percentage of subproblems that finish within $ \numparties $ iterations decreases with increasing numbers of patient-donor pairs, even for $ 40 $ pairs it still holds that only very few outliers require more than $ 3\numparties $ iterations. 

This empirically confirms that the general result that the Simplex algorithm terminates within $ 2m $ or $ 3m $ iterations for most practical problems also holds for kidney exchange using the real-world data set from UNOS. Thus, we conclude that $ \simplexFrequency = 2\numparties $ and $ \simplexFrequency = 3\numparties $ are good choices for obfuscating the information that can be derived from the number of Simplex iterations $ \numSimplexIterations $ without causing an unnecessarily high runtime overhead. However, the results from Figure~\ref{plt:boxplots_simplex_iterations} also indicate that the number of outliers that require more than $ 3\numparties $ Simplex iterations increases with increasing numbers of patient-donor pairs. If such outliers are not acceptable, it may be necessary to choose an even larger value for $ \simplexFrequency $.

\begin{table}[t]
	\caption{Number of subproblems required for solving the subset formulation of the KEP with our protocol~$ \kepIpProtocol $ for 10000 compatibility graphs where the patient-donor pairs are drawn uniformly at random from the UNOS data set (cf.~Section~\ref{sub:eval_setup}).}
	\label{tab:bab_interval_dist}
	\input{tables/bb_interval_dist.tex}
\end{table}

\customParagraph{Obfuscating the Structure of the Branch-and-Bound Tree.}
Although we did not find any obvious correlation between the structure of the Branch-and-Bound tree~$ \branchAndBoundStructure $ and any structural information, we still have evaluated the cost of obfuscating the information that can be derived from~$ \branchAndBoundStructure $. 
Similar to the approach for obfuscating the number of Simplex iterations, we just check for termination every $ \babFrequency \in \nats $ subproblems. 
Note that the protocol~$ \branchAndBoundProtocol $ (cf.~Protocol~\ref{prot:branch_and_bound}) is constructed such that once an optimal solution is found, this solution is not further modified by executing the protocol for additional subproblems. 

In contrast to the number of Simplex iterations, there is no general result known to date concerning the number of subproblems that the Branch-and-Bound algorithm requires for solving practical problems.
Therefore, we use the results of our empirical evaluation from Appendix~I to determine suitable values for the parameter $ \babFrequency $.

Table~\ref{tab:bab_interval_dist} shows the percentage of 10,000 instances of the KEP that finish with the indicated numbers of subproblems. We observe that most problem instances only require a single subproblem. While this percentage decreases with the number of patient-donor pairs, even for 40 pairs more than $ 90\% $ of all instances finish with at most three subproblems and less than $ 5\% $ of all instances require more than six~subproblems. 

Since a larger number of subproblems has a huge impact on the runtime performance of the protocol~$ \kepIpProtocol $ (e.g., increasing the number of subproblems from one to two, effectively doubles the runtime), we either consider $ \babFrequency = 1 $ or $ \babFrequency = 3 $ for our runtime measurements. However, the results shown in Table~\ref{tab:bab_interval_dist} also indicate that the percentages of instances that require more subproblems may further increase for larger numbers of patient-donor pairs. Thus, a larger value for $ \babFrequency $ may be necessary if the percentage of subproblems that require more than three subproblems becomes too large.

\begin{figure}[t]
	\includegraphics[width=\columnwidth]{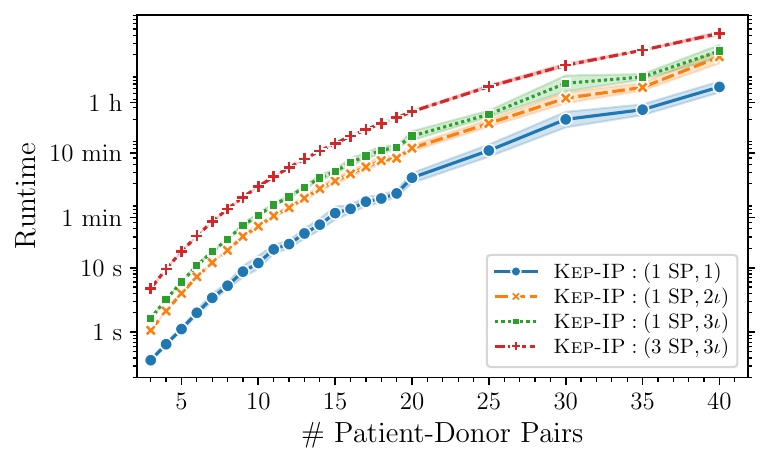}
	\vspace*{-2em}
	\caption{Runtime of the protocol~$ \textsc{Kep-IP} $ with different strategies for limiting the details on $ \numSimplexIterations $ and $ \branchAndBoundStructure $. A strategy $ (\babFrequency, \simplexFrequency) $ indicates that we check for termination at every $ \babFrequency $-th subproblem (SP) and at every $ \simplexFrequency $-th Simplex iteration.}
	\label{plt:runtimes_hide_simplex}
\end{figure}

\begin{table}[t]
	\caption{Runtime of the protocol~$ \textsc{Kep-IP} $ in seconds with four different strategies aimed to limit the information on $ \numSimplexIterations $ and $ \branchAndBoundStructure $. A strategy $ (\babFrequency, \simplexFrequency) $ indicates that we check for termination at every $ \babFrequency $-th subproblem (SP) and at every $ \simplexFrequency $-th Simplex iteration.}
	\label{tab:runtimes_hiding}
	\input{tables/runtimes_hiding.tex}
\end{table}

\customParagraph{Impact on the Runtime Performance.}
Based on the previously defined values for the parameters $ \simplexFrequency $ and $ \babFrequency $, we define different strategies for obfuscating the information on $ \numSimplexIterations $ and $ \branchAndBoundStructure $ in the output of the ideal functionality. A strategy $ (\babFrequency, \simplexFrequency) $ states that we check for termination at every $ \babFrequency $-th subproblem (SP) and at every $ \simplexFrequency $-th Simplex iteration. For example, a strategy $ (1 \ \text{SP}, 3\numparties) $ indicates that we check for termination at each subproblem but only for every $ 3\numparties $ Simplex iterations.

We evaluate our protocol~$ \kepIpProtocol $ for four different strategies. The first strategy corresponds to a regular execution of the protocol without any obfuscation of $ \numSimplexIterations $ and $ \branchAndBoundStructure $. Then, we consider two strategies where we still check for termination at each subproblem but only for every $ \simplexFrequency $-th Simplex iteration, once with $ \simplexFrequency = 2\numparties $ and once with $ \simplexFrequency = 3\numparties $. Finally, we consider the strategy where we only check for termination every third subproblem and every $ 3\numparties $ Simplex iterations. For each strategy, we execute $ 50 $ repetitions for each number of patient-donor pairs, drawing the pairs uniformly at random from the UNOS data set (cf.~Section~\ref{sub:eval_setup}).

Figure~\ref{plt:runtimes_hide_simplex} and Table~\ref{tab:runtimes_hiding} show the runtime differences of our protocol~$ \kepIpProtocol $ for different strategies for obfuscating the information on~$ \numSimplexIterations $ and $ \branchAndBoundStructure $. The evaluation setup is the same as in the run time evaluation for our protocol without implementing any obfuscation strategies on $ \numSimplexIterations $ and $ \branchAndBoundStructure $ (cf.~Section~\ref{sub:eval_setup}). In particular, we evaluate our protocol in MP-SPDZ~\cite{KellerMPSPDZ2020} using the secret sharing scheme from~\cite{Araki_ReplicatedSecretSharing_2016}. 
The line plots indicate the average runtime for each number of patient-donor pairs and the colored area around the line plots indicate the difference among the runtimes across different repetitions. 
Thus, the areas around the line plots also indicate in how far a strategy obfuscates the number of Simplex iterations~$ \numSimplexIterations $ and the structure of the Branch-and-Bound tree~$ \branchAndBoundStructure $. 

As expected, we observe that the runtime of the protocol increases with strategies that provide a higher level of obfuscation of $ \numSimplexIterations $ and $ \branchAndBoundStructure $. For example, for $ 20 $ patient-donor pairs the runtime on average increases by a factor of $ 2.9 $ if we only check for termination every $ 2\numparties $ Simplex iterations, by a factor of $ 4.4 $ if we check for termination every $ 3\numparties $ Simplex iterations, and by a factor of $ 10.6 $ if we check for termination every third subproblem and every $ 3\numparties $ Simplex iterations.

However, we also observe that there is less variance among the runtime of the different iterations with strategies that provide a higher level of obfuscation of $ \numSimplexIterations $ and $ \branchAndBoundStructure $.
This is indicated both by the smoother line plot and by the smaller colored area around the line plots. 
Note that the larger difference in the runtimes for strategy $ (1 \text{SP}, 3\numparties) $ compared to strategy $ (1 \text{SP}, 2\numparties) $ for some numbers of patient-donor pairs (e.g., for $ 35 $ pairs) stems from the fact that in many cases the number of Simplex iterations is already sufficiently obfuscated by the strategy $ (1 \text{SP}, 2\numparties) $. 
Since the strategy $ (1 \text{SP}, 3\numparties) $ further increases the number of Simplex iterations without obfuscating the number of subproblems, the difference in the runtime between instances that require a single subproblem and instances that require more than one subproblem increases even further. However, this does not indicate that the strategy is worse in obfuscating~$ \numSimplexIterations $ since in both cases the runtime difference only stems from the number of subproblems and not from the number of Simplex iterations.

In Appendix~I, we state that the inclusion of~$ \numSimplexIterations $ and $ \branchAndBoundStructure $ in the output of the ideal functionality allows for the deriving of some structural information, i.e., information on the underlying instance of the KEP that cannot be linked to particular patient-donor pairs (cf.~Section~\ref{sub:ideal_func}).
Our evaluations show that we can obfuscate the information that can be derived from~$ \numSimplexIterations $ and $ \branchAndBoundStructure $ to some degree, using different strategies that only check for termination at certain numbers of Simplex iterations and subproblems.
The choice of a particular strategy for obfuscating the information that can be derived from~$ \numSimplexIterations $ and $ \branchAndBoundStructure $ then always forms a trade-off between obfuscating more information on~$ \numSimplexIterations $ and $ \branchAndBoundStructure $ on the one hand and runtime performance on the other hand. 
For example, if statistics on the size of the optimal solution or other properties on the underlying compatibility graphs are desired, it may be acceptable to execute the protocol without any obfuscation of the Simplex iterations and the Branch-and-Bound tree. 
If such statistics are not desired, some strategy of obfuscation should be used.

\section*{Appendix III. Additional Evaluation Results}\label{app:plots}

\begin{figure*}
	\includegraphics[width=0.33\textwidth]{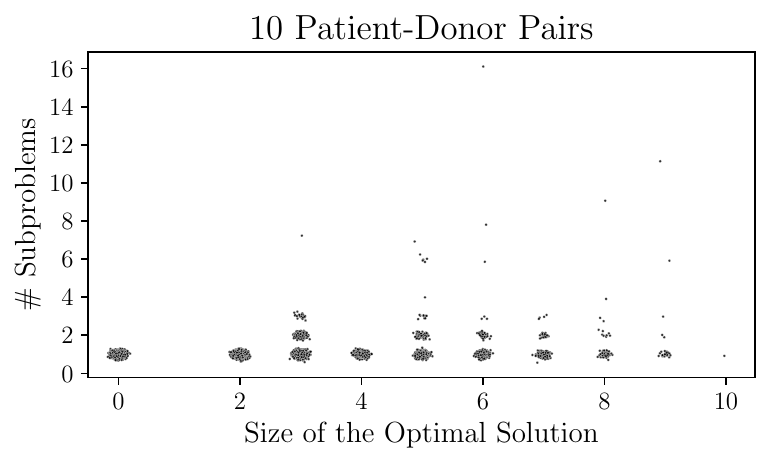}
	\includegraphics[width=0.33\textwidth]{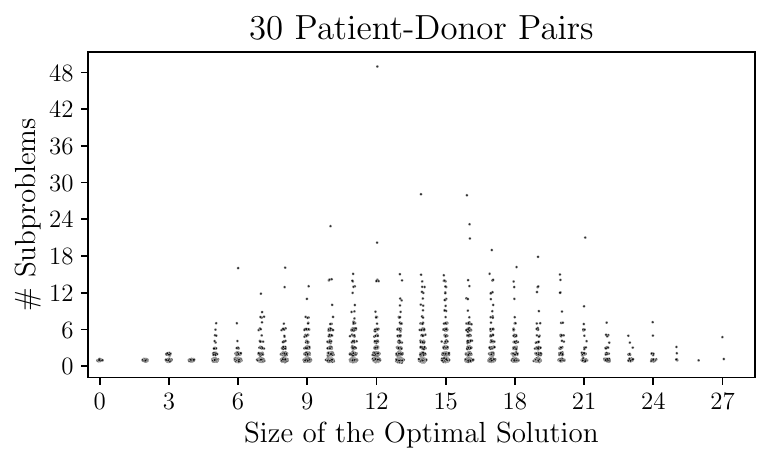}
	\includegraphics[width=0.33\textwidth]{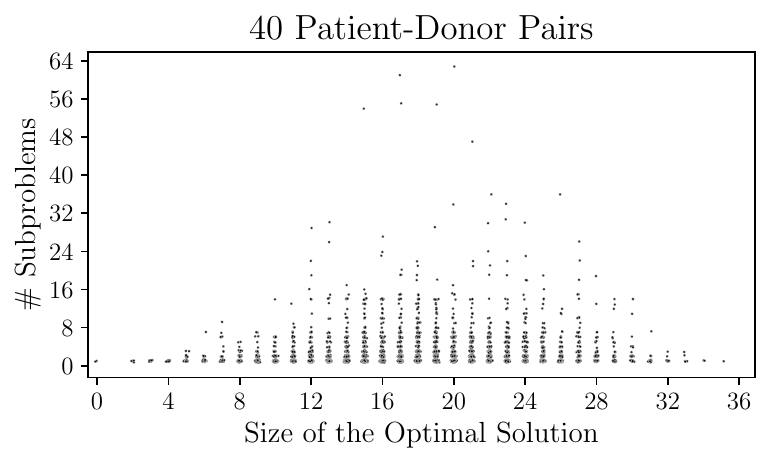}
	\vspace*{-2em}
	\caption{Relationship between the number of subproblems in the Branch-and-Bound tree for the protocol~$ \textsc{Kep-IP} $ and the size of the optimal solution for 10,000 instances of the KEP for different numbers of patient-donor pairs. Each dot corresponds to the number of subproblems for one execution of the Branch-and-Bound algorithm.}
	\label{plt:correlation_bab_plt_additional}
\end{figure*}

\begin{figure*}
	\includegraphics[width=0.5\textwidth]{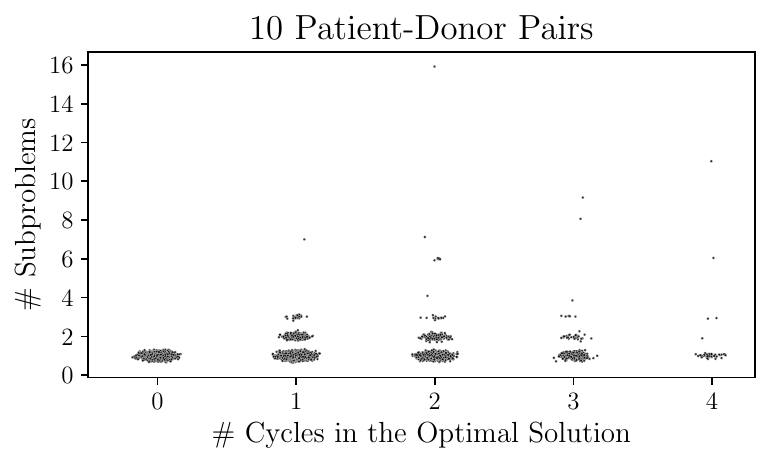}
	\includegraphics[width=0.5\textwidth]{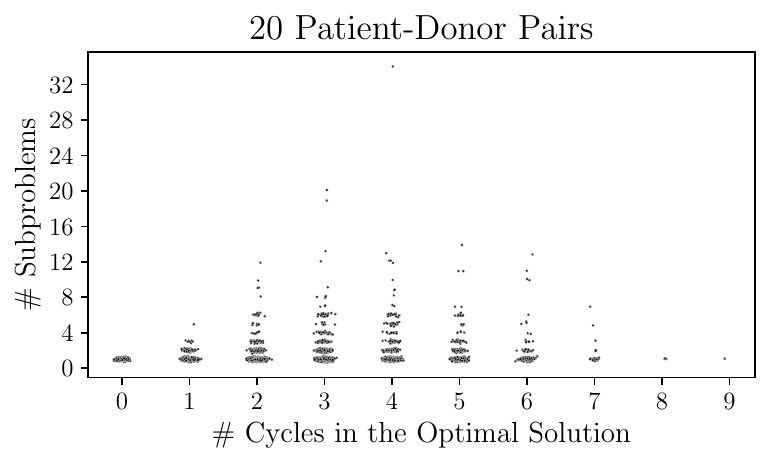}
	
	\includegraphics[width=0.5\textwidth]{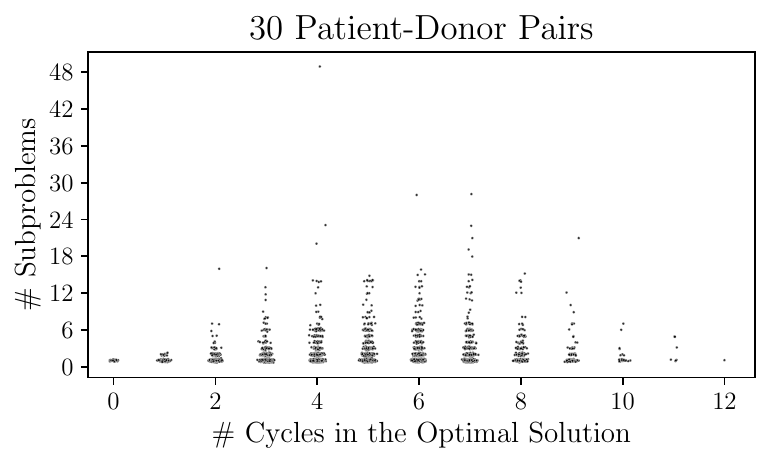}
	\includegraphics[width=0.5\textwidth]{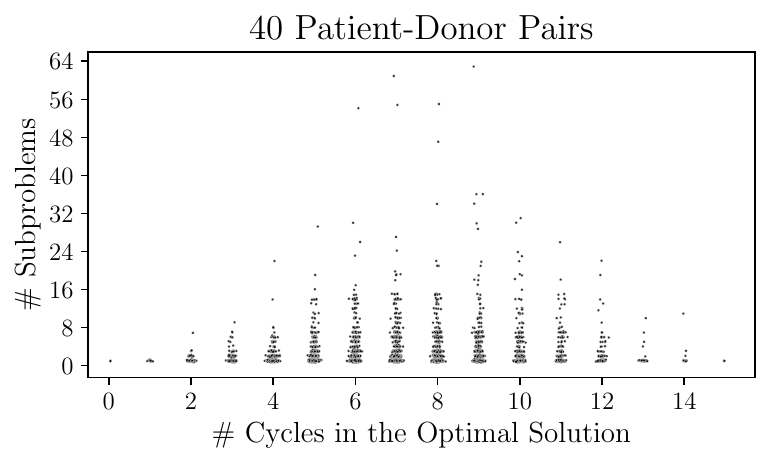}
	\vspace*{-2em}
	\caption{Relation between the number of subproblems in the Branch-and-Bound tree for the protocol~$ \textsc{Kep-IP} $ and the number of cycles in the computed optimal solution for 10,000 instances of the KEP for different numbers of patient-donor pairs. Each dot corresponds to the number of subproblems for one execution of the Branch-and-Bound algorithm.}
	\label{plt:correlation_bab_plt_comp_graph}
\end{figure*}

\begin{figure*}
	\includegraphics[width=0.5\textwidth]{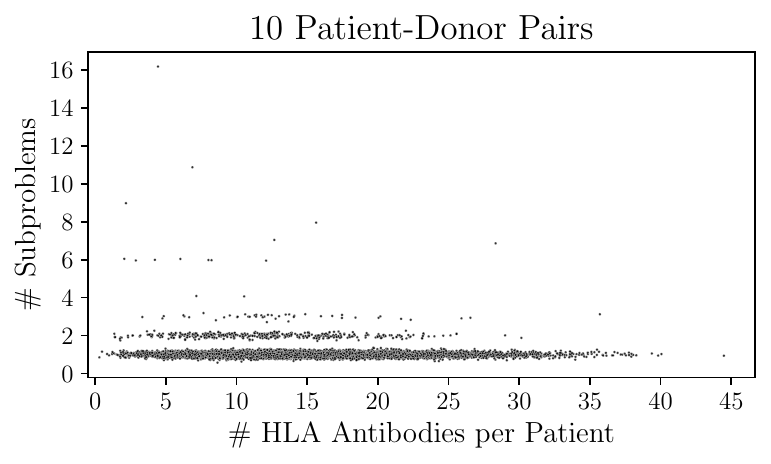}
	\includegraphics[width=0.5\textwidth]{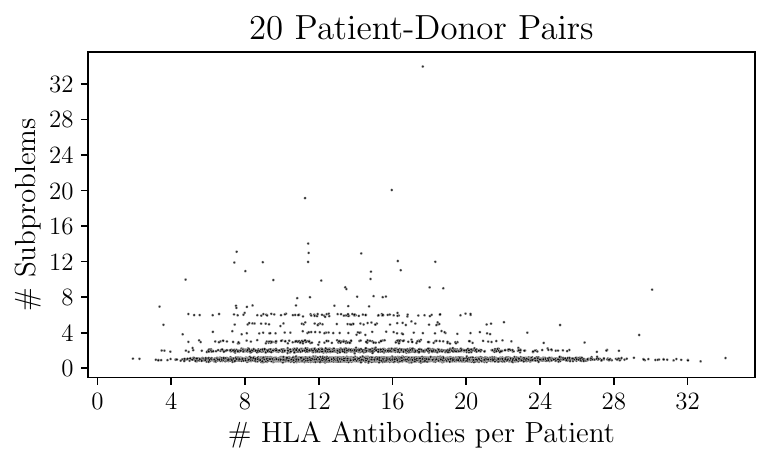}
	
	\includegraphics[width=0.5\textwidth]{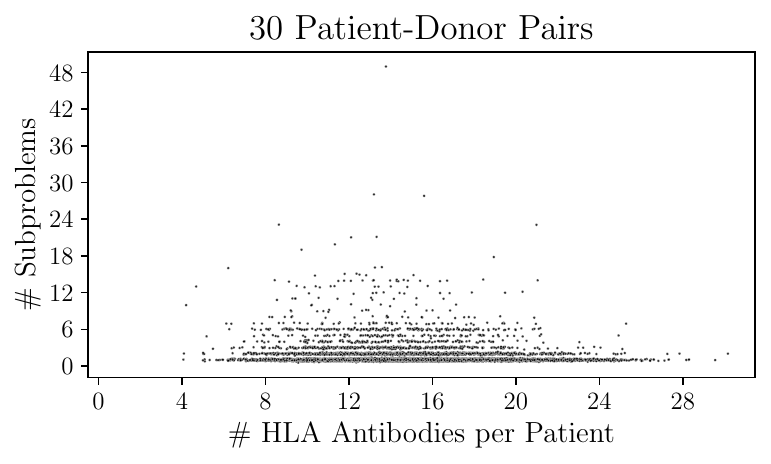}
	\includegraphics[width=0.5\textwidth]{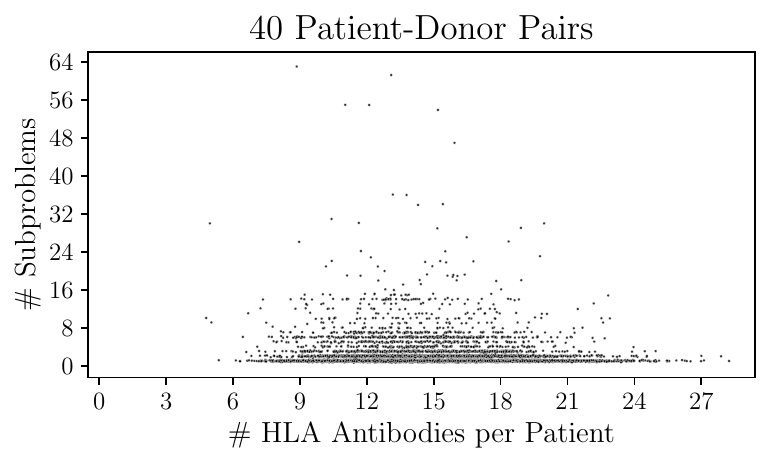}
	\vspace*{-2em}
	\caption{Relationship between the number of subproblems in the Branch-and-Bound tree for the protocol~$ \textsc{Kep-IP} $ and the average number of HLA antibodies of the patients for 10,000 instances of the KEP for different numbers of patient-donor pairs. Each dot corresponds to the number of subproblems for one execution of the Branch-and-Bound algorithm.}
	\label{plt:correlation_bab_plt_antibodies}
\end{figure*}

In this section, we provide additional evaluation results for the analysis presented in Appendix~I. Figure~\ref{plt:correlation_simplex_plt_additional} shows the relationship between the number of Simplex iterations and the number of cycles in the optimal solution for additional numbers of patient-donor pairs.  
Figure~\ref{plt:correlation_bab_plt_additional} shows the relationship between the number of subproblems in the Branch-and-Bound tree and the size of the optimal solution for additional numbers of patient-donor pairs.  
Figure~\ref{plt:correlation_bab_plt_comp_graph} and Figure~\ref{plt:correlation_bab_plt_antibodies} show the relationship between the number of subproblems in the Branch-and-Bound tree and the number of cycles that exist in the optimal solution and the average number of antibodies per patient, respectively.

\begin{figure*}[t]
	\includegraphics[width=0.33\textwidth]{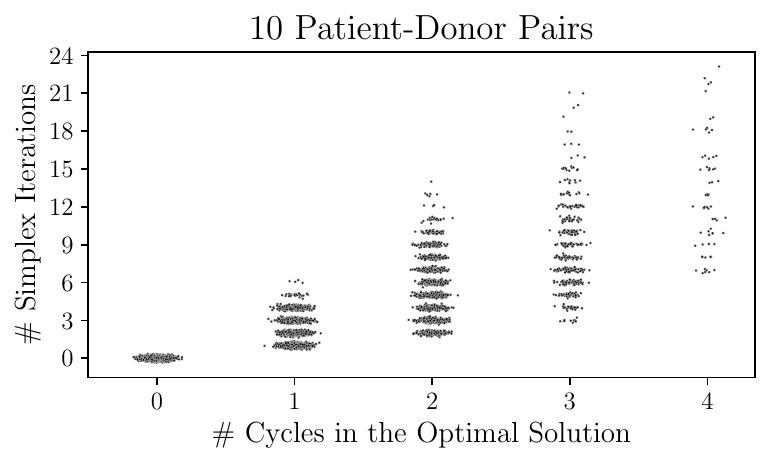}
	\includegraphics[width=0.33\textwidth]{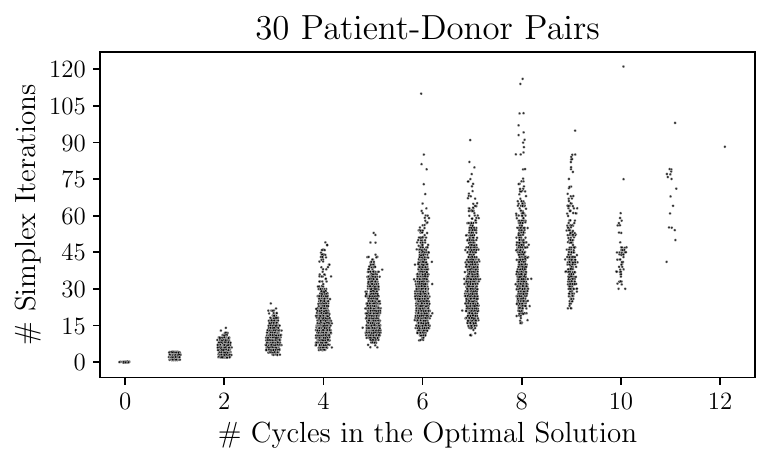}
	\includegraphics[width=0.33\textwidth]{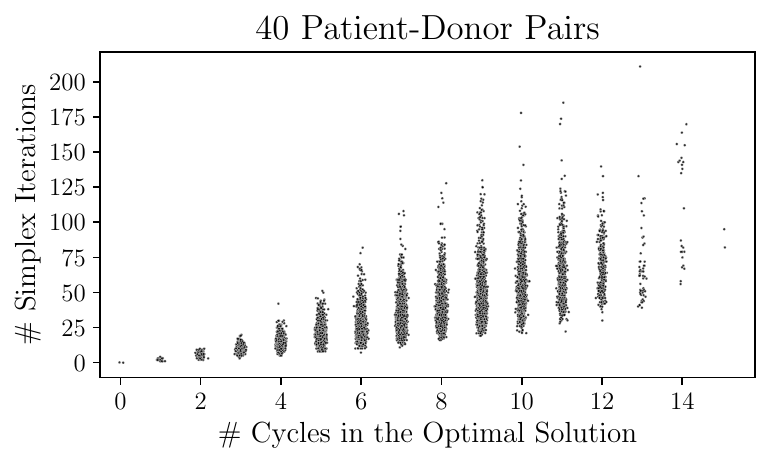}
	\vspace*{-2em}
	\caption{Relationship between the number of Simplex iterations for each node in the Branch-and-Bound tree $ \branchAndBoundStructure $ and the number of cycles that are included in the corresponding computed optimal solution for 10,000 instances of the KEP for different numbers of patient-donor pairs. Each dot corresponds to the number of Simplex iterations for one execution of the Simplex algorithm.}
	\label{plt:correlation_simplex_plt_additional}
\end{figure*}

%% file: tables/bb_interval_dist.tex
\setlength{\tabcolsep}{3pt}\centering
\begin{tabular}{c | c c c c c c c}
\multicolumn{1}{c}{} & \multicolumn{7}{c}{\# subproblems} \Bstrut\\
$ \numparties $ & 1 & 2 & 3 & 4-6 & 7-10 & 11-20 & 21-126\Tstrut\Bstrut\\\hline
5 & 99.19\% & \phantom{0}0.75\% & \phantom{0}0.06\% & \phantom{0}0.00\% & \phantom{0}0.00\% & \phantom{0}0.00\% & \phantom{0}0.00\% \Tstrut\\
10 & 95.53\% & \phantom{0}3.88\% & \phantom{0}0.44\% & \phantom{0}0.09\% & \phantom{0}0.04\% & \phantom{0}0.02\% & \phantom{0}0.00\% \Tstrut\\
15 & 91.03\% & \phantom{0}7.40\% & \phantom{0}0.81\% & \phantom{0}0.62\% & \phantom{0}0.07\% & \phantom{0}0.06\% & \phantom{0}0.01\% \Tstrut\\
20 & 86.18\% & 10.24\% & \phantom{0}1.48\% & \phantom{0}1.71\% & \phantom{0}0.24\% & \phantom{0}0.14\% & \phantom{0}0.01\% \Tstrut\\
25 & 81.83\% & 12.01\% & \phantom{0}2.47\% & \phantom{0}2.87\% & \phantom{0}0.45\% & \phantom{0}0.31\% & \phantom{0}0.06\% \Tstrut\\
30 & 77.68\% & 13.10\% & \phantom{0}3.54\% & \phantom{0}4.00\% & \phantom{0}0.95\% & \phantom{0}0.66\% & \phantom{0}0.07\% \Tstrut\\
35 & 73.48\% & 14.53\% & \phantom{0}4.17\% & \phantom{0}5.35\% & \phantom{0}1.38\% & \phantom{0}0.88\% & \phantom{0}0.21\% \Tstrut\\
40 & 70.76\% & 14.32\% & \phantom{0}5.11\% & \phantom{0}5.84\% & \phantom{0}2.22\% & \phantom{0}1.44\% & \phantom{0}0.31\% \Tstrut\\
\end{tabular}

%% file: tables/runtimes_hiding.tex
\centering
\begin{tabular}{c | c c c c}
$\numparties$ & $(1\text{ SP}, 1)$ & $(1\text{ SP}, 2\iota)$ & $(1\text{ SP}, 3\iota)$ & $(3\text{ SP}, 3\iota)$ \Bstrut\\\hline
5 & \phantom{0000}1 & \phantom{0000}4 & \phantom{0000}6 & \phantom{000}18 \Tstrut\Bstrut\\
10 & \phantom{000}12 & \phantom{000}44 & \phantom{000}65 & \phantom{00}181\Bstrut\\
15 & \phantom{000}70 & \phantom{00}220 & \phantom{00}309 & \phantom{00}838\Bstrut\\
20 & \phantom{00}247 & \phantom{00}710 & \phantom{0}1091 & \phantom{0}2617\Bstrut\\
25 & \phantom{00}653 & \phantom{0}1720 & \phantom{0}2379 & \phantom{0}6391\Bstrut\\
30 & \phantom{0}1983 & \phantom{0}4227 & \phantom{0}7182 & 13692\Bstrut\\
35 & \phantom{0}2802 & \phantom{0}6198 & \phantom{0}9001 & 23399\Bstrut\\
40 & \phantom{0}6303 & 18532 & 22610 & 42580\Bstrut\\
\end{tabular}

%% file: conference_101719.bbl
\begin{thebibliography}{10}
\providecommand{\url}[1]{#1}
\csname url@samestyle\endcsname
\providecommand{\newblock}{\relax}
\providecommand{\bibinfo}[2]{#2}
\providecommand{\BIBentrySTDinterwordspacing}{\spaceskip=0pt\relax}
\providecommand{\BIBentryALTinterwordstretchfactor}{4}
\providecommand{\BIBentryALTinterwordspacing}{\spaceskip=\fontdimen2\font plus
\BIBentryALTinterwordstretchfactor\fontdimen3\font minus
  \fontdimen4\font\relax}
\providecommand{\BIBforeignlanguage}[2]{{%
\expandafter\ifx\csname l@#1\endcsname\relax
\typeout{** WARNING: IEEEtranS.bst: No hyphenation pattern has been}%
\typeout{** loaded for the language `#1'. Using the pattern for}%
\typeout{** the default language instead.}%
\else
\language=\csname l@#1\endcsname
\fi
#2}}
\providecommand{\BIBdecl}{\relax}
\BIBdecl

\bibitem{AbrahamClearingAlgorithms2007}
D.~J. Abraham, A.~Blum, and T.~Sandholm, ``Clearing algorithms for barter
  exchange markets: Enabling nationwide kidney exchange,'' \emph{ACM Conference
  on Electronic Commerce}, 2007.

\bibitem{Araki_ReplicatedSecretSharing_2016}
T.~Araki, J.~Furukawa, Y.~Lindell, A.~Nof, and K.~Ohara, ``High-throughput
  semi-honest secure three-party computation with an honest majority,'' in
  \emph{Proceedings of the 2016 ACM SIGSAC Conference on Computer and
  Communications Security}.\hskip 1em plus 0.5em minus 0.4em\relax ACM, 2016.

\bibitem{Araki_SecureGraphAnalysis_2021}
T.~Araki, J.~Furukawa, K.~Ohara, B.~Pinkas, H.~Rosemarin, and H.~Tsuchida,
  ``Secure graph analysis at scale,'' in \emph{Proceedings of the 2021 ACM
  SIGSAC Conference on Computer and Communications Security}, 2021.

\bibitem{AshlagiKidneyExchangeOperations2021}
I.~Ashlagi and A.~E. Roth, ``Kidney exchange: An operations perspective,''
  National Bureau of Economic Research, Tech. Rep., 2021.

\bibitem{Birka_PPKidneyExchange_2022}
T.~Birka, K.~Hamacher, T.~Kussel, H.~M{\"o}llering, and T.~Schneider, ``Spike:
  Secure and private investigation of the kidney exchange problem,'' \emph{BMC
  Medical Informatics and Decision Making}, vol.~22, no.~1, 2022.

\bibitem{Biro_EuropeanKE_2019}
P.~Bir{\'o}, B.~Haase-Kromwijk, T.~Andersson, E.~I. {\'A}sgeirsson,
  T.~Baltesov{\'a}, I.~Boletis, C.~Bolotinha, G.~Bond, G.~B{\"o}hmig,
  L.~Burnapp \emph{et~al.}, ``Building kidney exchange programmes in europe --
  an overview of exchange practice and activities,'' \emph{Transplantation},
  vol. 103, no.~7, 2019.

\bibitem{Breuer_KepIp_2022}
M.~Breuer, P.~Hein, L.~Pompe, B.~Temme, U.~Meyer, and S.~Wetzel, ``Solving the
  kidney exchange problem using privacy-preserving integer programming,'' in
  \emph{2022 19th Annual International Conference on Privacy, Security \& Trust
  (PST)}.\hskip 1em plus 0.5em minus 0.4em\relax IEEE, 2022.

\bibitem{Breuer_Matching_2022}
M.~Breuer, U.~Meyer, and S.~Wetzel, ``Privacy-preserving maximum matching on
  general graphs and its application to enable privacy-preserving kidney
  exchange,'' in \emph{Twelveth ACM Conference on Data and Application Security
  and Privacy}.\hskip 1em plus 0.5em minus 0.4em\relax ACM, 2022.

\bibitem{Breuer_KEprotocol_2020}
M.~Breuer, U.~Meyer, S.~Wetzel, and A.~M{\"u}hlfeld, ``A privacy-preserving
  protocol for the kidney exchange problem,'' in \emph{Workshop on Privacy in
  the Electronic Society}.\hskip 1em plus 0.5em minus 0.4em\relax ACM, 2020.

\bibitem{Catrina_PrimitivesSMPC_2010}
O.~Catrina and S.~De~Hoogh, ``Improved primitives for secure multiparty integer
  computation,'' in \emph{International Conference on Security and Cryptography
  for Networks}.\hskip 1em plus 0.5em minus 0.4em\relax Springer, 2010.

\bibitem{Catrina_PPLinearProgramming_2010}
O.~Catrina and S.~d. Hoogh, ``Secure multiparty linear programming using
  fixed-point arithmetic,'' in \emph{European Symposium on Research in Computer
  Security}.\hskip 1em plus 0.5em minus 0.4em\relax Springer, 2010.

\bibitem{Damgard_ABB_2003}
I.~Damg{\aa}rd and J.~B. Nielsen, ``Universally composable efficient multiparty
  computation from threshold homomorphic encryption,'' in \emph{Annual
  International Cryptology Conference}.\hskip 1em plus 0.5em minus 0.4em\relax
  Springer, 2003.

\bibitem{Dantzig_LinearProgrammingSimplex_1963}
G.~Dantzig, \emph{Linear Programming and Extensions}.\hskip 1em plus 0.5em
  minus 0.4em\relax Princeton University Press, 1963.

\bibitem{Delorme_KEGraphGenerator_2022}
M.~Delorme, S.~Garc{\'\i}a, J.~Gondzio, J.~Kalcsics, D.~Manlove, W.~Pettersson,
  and J.~Trimble, ``Improved instance generation for kidney exchange
  programmes,'' \emph{Computers \& Operations Research}, vol. 141, 2022.

\bibitem{Dreier_PPLinearProgramming_2011}
J.~Dreier and F.~Kerschbaum, ``Practical secure and efficient multiparty linear
  programming based on problem transformation,'' in \emph{IACR Cryptology
  ePrint Archive}, 2011.

\bibitem{Escudero_edaBits_2020}
D.~Escudero, S.~Ghosh, M.~Keller, R.~Rachuri, and P.~Scholl, ``Improved
  primitives for mpc over mixed arithmetic-binary circuits,'' in \emph{Annual
  International Cryptology conference}.\hskip 1em plus 0.5em minus 0.4em\relax
  Springer, 2020.

\bibitem{GoldreichFoundationsTwo2004}
O.~Goldreich, \emph{Foundations of Cryptography: Volume 2 - Basic
  Applications}.\hskip 1em plus 0.5em minus 0.4em\relax Cambridge University
  Press, 2004.

\bibitem{Grishpoun_PPDCOP_2016}
T.~Grinshpoun and T.~Tassa, ``P-syncbb: A privacy preserving branch and bound
  dcop algorithm,'' \emph{Journal of Artificial Intelligence Research},
  vol.~57, 2016.

\bibitem{KellerMPSPDZ2020}
M.~Keller, ``{MP-SPDZ}: A versatile framework for multi-party computation,'' in
  \emph{Computer and Communications Security}.\hskip 1em plus 0.5em minus
  0.4em\relax ACM, 2020.

\bibitem{Land_BranchAndBound_1960}
A.~H. Land and A.~G. Doig, ``An automatic method for solving discrete
  programming problems,'' \emph{Econometrica}, vol.~28, no.~3, 1960.

\bibitem{Li_PPLinearProgramming}
J.~Li and M.~J. Atallah, ``Secure and private collaborative linear
  programming,'' in \emph{International Conference on Collaborative Computing:
  Networking, Applications and Worksharing}.\hskip 1em plus 0.5em minus
  0.4em\relax IEEE, 2006.

\bibitem{Nocedal_NumericalOptimization_1999}
J.~Nocedal and S.~J. Wright, \emph{Numerical optimization}.\hskip 1em plus
  0.5em minus 0.4em\relax Springer, 1999.

\bibitem{Quandt_SimplexAverageTermination_1964}
R.~E. Quandt and H.~W. Kuhn, ``On upper bounds for the number of iterations in
  solving linear programs,'' \emph{Operations Research}, vol.~12, no.~1, 1964.

\bibitem{Saidman_KEGraphGenerator_2006}
S.~L. Saidman, A.~E. Roth, T.~S{\"o}nmez, M.~U. {\"U}nver, and F.~L. Delmonico,
  ``Increasing the opportunity of live kidney donation by matching for two-and
  three-way exchanges,'' \emph{Transplantation}, vol.~81, no.~5, 2006.

\bibitem{Schrijver_LinearProgramming_1998}
A.~Schrijver, \emph{Theory of linear and integer programming}.\hskip 1em plus
  0.5em minus 0.4em\relax John Wiley \& Sons, 1998.

\bibitem{ShamirSecretSharing1979}
A.~Shamir, ``How to share a secret,'' \emph{Communications of the ACM},
  vol.~22, no.~11, 1979.

\bibitem{Tassa_PPDCOP_2021}
T.~Tassa, T.~Grinshpoun, and A.~Yanai, ``Pc-syncbb: A privacy preserving
  collusion secure dcop algorithm,'' \emph{Artificial Intelligence}, vol. 297,
  2021.

\bibitem{Toft_PPLinearProgramming_2009}
T.~Toft, ``Solving linear programs using multiparty computation,'' in
  \emph{Financial Cryptography and Data Security}.\hskip 1em plus 0.5em minus
  0.4em\relax Springer, 2009.

\bibitem{Vaidya_PPLinearProgramming_2009}
J.~Vaidya, ``Privacy-preserving linear programming,'' in \emph{ACM Symposium on
  Applied Computing}.\hskip 1em plus 0.5em minus 0.4em\relax ACM, 2009.

\bibitem{Zahur_shuffling_2016}
S.~Zahur, X.~Wang, M.~Raykova, A.~Gasc{\'o}n, J.~Doerner, D.~Evans, and
  J.~Katz, ``Revisiting square-root oram: efficient random access in
  multi-party computation,'' in \emph{2016 IEEE Symposium on Security and
  Privacy (SP)}.\hskip 1em plus 0.5em minus 0.4em\relax IEEE, 2016.

\end{thebibliography}
